\documentclass[apj]{emulateapj}

\usepackage{apjfonts}
\usepackage{times}

\usepackage{amsmath}
\usepackage{graphicx}

\usepackage{natbib}

\usepackage{color}

\slugcomment{Received 2010 May 11; accepted 2019 Nov 10}
\shorttitle{}
\shortauthors{Le \& Woo}

\newcommand{\Hb}{H{$\beta$}}

\newcommand{\FeII}{\ion{Fe}{2}}

\newcommand{\MgII}{\ion{Mg}{2}}

\newcommand{\FeUV}{\ion{Fe}{2}$_{\rm uv}$} 
\newcommand{\FeOPT}{\ion{Fe}{2}$_{\rm  opt}$} 

\newcommand{\OIII}{[\ion{O}{3}]}
\newcommand{\OII}{[\ion{O}{2}]}

\newcommand{\mbh}{M$_{\rm BH}$}

\newcommand{\kms}{km~s$^{\rm -1}$}
\newcommand{\ergs}{erg s$^{-1}$}

\begin{document}

\title{Comparison of The UV and Optical \FeII\ Emission in Type 1 AGNs}



\author{Huynh Anh N. Le$^{1,2,3}$}
\author{Jong-Hak Woo$^{1}$}
\affil{$^{1}$Astronomy Program, Department of Physics and Astronomy, Seoul National University, Seoul, 08826; woo@astro.snu.ac.kr\\
$^{2}$CAS Key Laboratory for Research in Galaxies and Cosmology, Department of Astronomy, University of Science and Technology of China, Hefei 230026, China; lha@ustc.edu.cn\\
$^{3}$School of Astronomy and Space Science, University of Science and Technology of China, Hefei 230026, China
}
%

\begin{abstract}

We present the kinematical properties of the UV and optical \FeII\ emission gas based on the velocity shift and line width 
measurements of a sample of 223 Type 1 active galactic nuclei (AGNs) at 0.4 $<$ z $<$ 0.8. We find a strong correlation between the line widths of the UV and optical \FeII\ emission lines, indicating that both \FeII\ emission features arise from similar distances in the broad line region (BLR). However, in detail we find differing trends, depending on the width of \FeII. While the velocity shifts and dispersions of the UV Fe II (\FeUV) and
optical Fe II (\FeOPT) emission lines are comparable to each other for AGNs with relatively narrow \FeOPT\ line widths (i.e., FWHM < 3200 \kms;
Group A), \FeOPT\ is broader than \FeUV\ for AGNs with relatively broad \FeOPT\  (i.e., FWHM > 3200 \kms; Group B). \FeII\ emission lines are on average narrower than \Hb\ and \MgII\ for Group A, indicating the \FeII\ emission region is further out in the BLR, while for Group B AGNs \FeOPT\ is comparable to \Hb\ and broader than \MgII.
While \FeII\ emission lines are on average redshifted ($40\pm141$ \kms\ and $182\pm95$, respectively for \FeUV\ and \FeOPT),
indicating inflow, the sample as a whole shows a large range of velocity shifts, suggesting complex nature of gas kinematics.


\end{abstract}

\keywords{galaxies: active -- galaxies: nuclei -- quasars: emission lines}

\section{INTRODUCTION} \label{section:intro}

Active galactic nuclei (AGNs) present diverse components of broad and narrow emission lines in the UV to optical spectral range. These spectral properties reflects the geometric structure, distribution of emissivity, and kinematics of the gas. The \FeII\ emission blends are often strongly detected around the UV \MgII\ 2798\AA\ and the optical \Hb\ 4861\AA\ lines. Since 1960s, a number of studies has investigated the properties of \FeII\ emission (e.g., \citealp{Greenstein&Schmidt64, Kwan&Krolik81, Joly87}; \citealp{Wampler&Oke67}). 
Iron is mainly produced by Type Ia supernovae (SNe) in a binary system, and by Type II SNe in massive stars along with heavy $\alpha$ elements, e.g., O, Ne, and Mg. Because of their different time scales of formation, the \FeII/\MgII\ line flux ratio is often used to investigate chemical evolution as the first order-proxy for the Fe/Mg element ratio \citep[e.g,][]{DeRosa+11, Sameshima+17, Shin+19}.  

The studies of the excitation mechanism and correlation with other emission line properties are important to understanding the origin of \FeII\ emission, as there are a number of debates on the nature and origin of \FeII\ in the literature. \FeII\ emission was considered to originate from the same region as broad emission lines, i.e., \Hb. For example, using 87 low-z quasars, for example, \citet{BG92} reported that the line widths of \FeII\ and \Hb\ lines were comparable, indicting that they were at similar locations in the broad line region (BLR).
In a recent study, however, \citet{Popovic+07} reported that the full-width-at-half-maximum (FWHM) of \FeII\ is $\frac{1}{3}$ of that of the broad \Hb\ line. Moreover, \citet{Hu+08a} showed that the FWHM of \FeII\ line is narrower than that of \Hb\ (i.e., FWHM \FeII\ $\sim$ $\frac{3}{4}$ FWHM of \Hb). These results indicate that \FeII\ emission may originate from a region that is located further out in the BLR, compared to that emitting the bulk of \Hb.

The velocity shift of \FeII\ emission is also a source of debate. A number of previous studies assumed no systemic velocity shift of \FeII\ with respect to \OII\ (e.g., \citealp{BG92}; \citealp{Kim+06}; \citealp{Greene&Ho05}). However, \citet{Hu+08} reported that the optical \FeII\ emission is redshifted by $\sim$400$-$2000 \kms\ with respect to the peak of the \OIII\ 5007\AA\ line, using a set of composite spectra, which were constructed with the AGNs at z $<$ 0.8 selected from the Sloan Digital Sky Survey (SDSS). In contrast, \citet{Sulentic+12} disagreed to with \citet{Hu+08}, arguing that the measurement of \FeII\ velocity shifts is reliable only for AGNs with high signal-to-noise ratio (S/N) spectra. Based on their high S/N composite spectra, \citet{Sulentic+12} claimed no redshift of the optical \FeII\ emission. Later on, \citet{Kovacevic+15} reported no significant redshift of the optical \FeII\ emission using a sample of 293 SDSS AGNs at 0.4 $<$ z $<$ 0.6, while they showed on average a large redshift of the UV \FeII\ emission with respect to a narrow emission line [OIII] 5007\AA\ \citep[see also][]{Kovacevic+10}. For the origin of the difference between \FeUV\ and \FeOPT, \citet{Kovacevic+15} suggested the possibility of the asymmetric distribution of the UV-emitting gas clouds in the BLR, as well as the effect of internal shock increasing excitation of UV lines in infalling gas. 


There are also issues in the models of the UV and optical \FeII\ emission, since classical photoionization models cannot explain the measured flux ratio between the optical and UV \FeII\ emission (e.g., \citealp{Collin&Joly00}; \citealp{Baldwin+04}; \citealp{Sameshima+11}). 
To solve this problem, for example, \citet{Baldwin+04} discussed local micro-turbulence in the emitting gas in their model. \citet{Ferland+09} investigated 
the role of anisotropy in \FeII\ emission from local clouds as the anisotropy depends on the column density of individual clouds, potentially causing 
the difference between optical and UV \FeII\ emission.



Overall, studying the physical properties of \FeII\ emission is intriguing. Multiple \FeII\ emission lines are blended in a typical AGN spectrum, which is composed of the continuum from the central source (accretion disk and reprocessed photons), and host galaxy contribution. Determining accurate Balmer continuum is also difficult since the best-fit result depends on the assumed physical parameters, i.e., electron temperature $\rm T_{e}$, electron density $\rm n_{e}$, optical depths, the shape of the continuum, and the UV \FeII\ models (e.g., \citealp{Kovacevic+14} and \citealp{Kovacevic+17}). Clearly, high S/N spectra and careful spectral decomposition are required to archive reliable measurements of \FeII\ emission properties.

In this paper, we investigate the kinematical properties of the UV and optical \FeII\ emission, using a sample of 223 AGNs at 0.4 $<$ z $<$ 0.8. The sample is a composed of
38 moderate-luminosity AGNs with high quality Keck spectra (S/N $\geq$ 10 in 3000\AA\ and $\geq$ 20 in 5100\AA) 
from our previous study \citep{Woo+18}, and 185 AGNs with S/N $\geqslant$ 20 in both 3000 \AA\ and 5100 \AA\ selected from SDSS. The subsamples obtained from SDSS and Keck are complementary to each other, enabling us to explore the physical connections between the UV and optical \FeII\ emission over a large range of AGN luminosity.
Using these AGNs, we investigate \FeII\ velocity shift, using individual spectra rather than using a composite spectra as previously done by \citet{Sulentic+12} and \citet{Hu+12}. One of our main goals  is to settle the debate between \citet{Hu+08} and \citet{Sulentic+12} on the systemic redshift of the optical \FeII\ emission lines. We present the comparison of the UV and optical \FeII\ line properties along with the kinematical properties of the broad emission lines, i.e., \Hb\ and \MgII. 
Section 2 describes the sample selection, and Section 3 presents the method. The main results are presented in Section 4, followed by Discussion and summary in Section 5 and 6, respectively. The following cosmological parameters are used throughout the paper: $H_0 = 70$~km~s$^{-1}$~Mpc$^{-1}$, $\Omega_{\rm m} = 0.30$, and $\Omega_{\Lambda} = 0.70$.

\section{Sample Selection}\label{section:obs}

We selected a sample of 52 moderate-luminosity AGNs ($\mathrm{\lambda L_{5100} \sim10^{43.8} - 10^{44.4}}$ \ergs) at redshift of 0.4 $<$ z $<$ 0.6, which was initially chosen for measuring stellar velocity dispersions of AGN host galaxies to study the evolution of the black hole mass and host spheroid velocity dispersion ($\rm M_{\rm BH} - \sigma_{*}$) relation \citep{Treu04, Woo06, Woo+08, Bennert+10, Park+15}, and was also presented in a series of papers in the study of the UV and optical \mbh\ estimators (\citealp{McGill+08} and \citealp{Woo+18}). Observations and data reduction process of the sample can be found in \citet{Park+15, Woo+18} for the optical and UV spectral ranges. 
Among 52 targets, we removed 5 AGNs with strong internal extinction (see details in \citealp{Woo+18}). Among the remaining 47 targets, we selected AGNs based on a couple of criteria: first, S/N $>$ 10 in the continuum at 3000 \AA\ and 20 at 5100 \AA. Second, the UV and optical \FeII\ emissions are strong, i.e., the contribution of the \FeII\ emission is larger than 20\% of the total flux in the UV spectral range of 2600 $-$ 3050 \AA, and in the optical spectral range of 4434 $-$ 4684 \AA, respectively. As a result, we obtained 38 AGNs with high quality spectra.

We also used the Sloan Digital Sky Survey (SDSS) quasar catalog \citep{Shen+11} to initially select 14,367 quasars at 0.4 $<$ z $<$ 0.8. The choice of the redshift range is to include both the rest-frame UV and optical \FeII\ lines. Then, we chose 487 AGNs with high quality spectra (i.e., S/N $\geqslant$ 20) at both 3000 \AA\ and 5100 \AA. Among these AGNs, we removed 63 targets, which contain strong absorption lines in the \MgII\ line profile. In addition, we excluded 10 targets, which have relatively weak \OIII\ lines and poor fitting results. From the remaining 414 targets, we selected 321 AGNs, which have the contribution of the \FeII\ emission is greater than 20\% in the pseudo continuum as applied to the Keck sample.

After analyzing all emission lines, we further removed the targets with large fractional errors ($\geqslant$ 50\%) of the line width and velocity shift of \FeII\ emission lines (see Section 3.3). The final SDSS sample contained 185 targets with the monochromatic luminosity at 5100\AA\ $\lambda L_{5100}$ ranging from 10$^{44.5}$ to 10$^{46.5}$ \ergs. By combining the SDSS AGNs with the moderate luminosity AGNs from our previous study, we enlarged the luminosity range of the sample for comparing the UV and optical \FeII\ emission line kinematics.

\begin{figure*}
\figurenum{1}
\center
	\includegraphics[width = 0.4\textwidth]{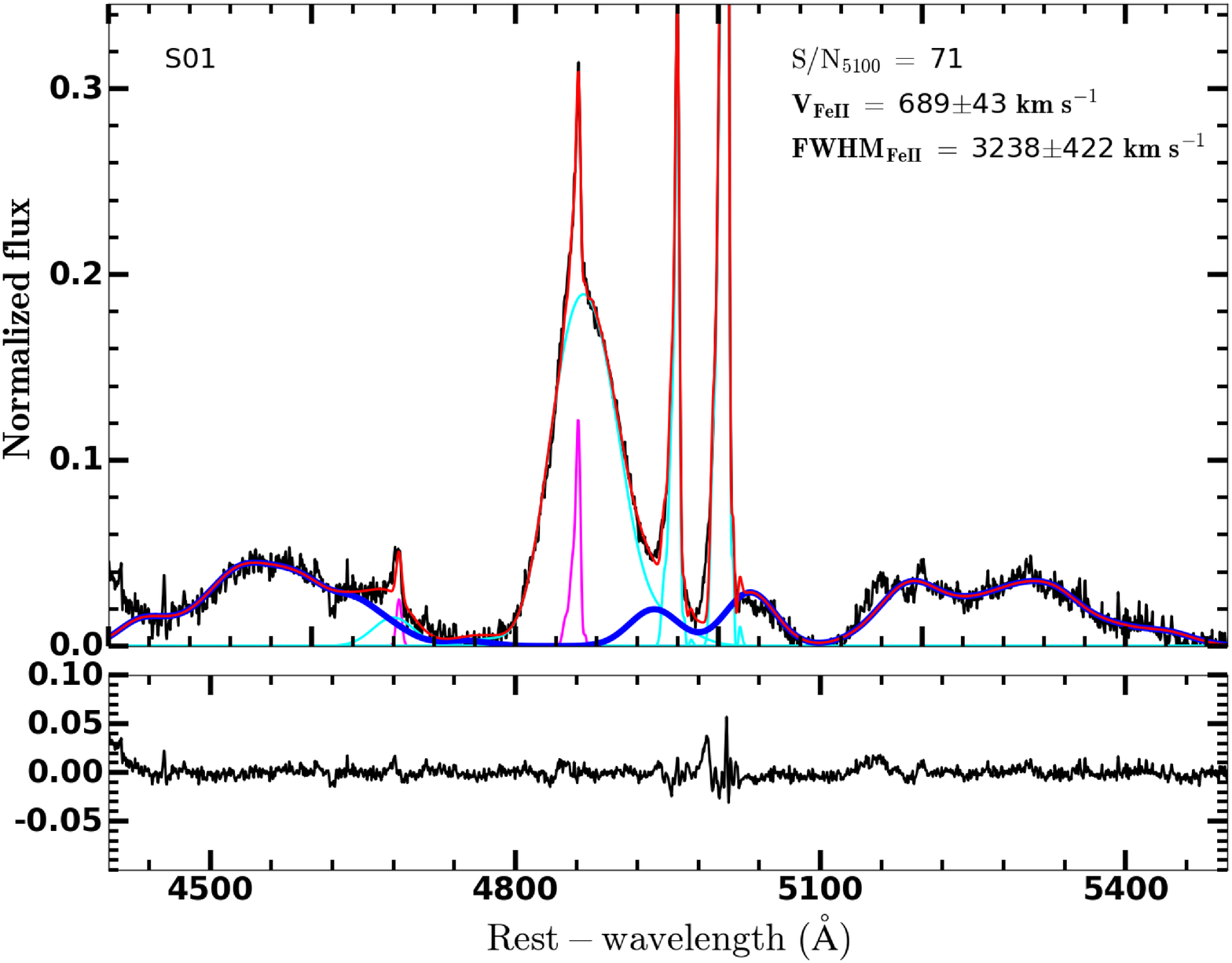}
	\includegraphics[width = 0.315\textwidth]{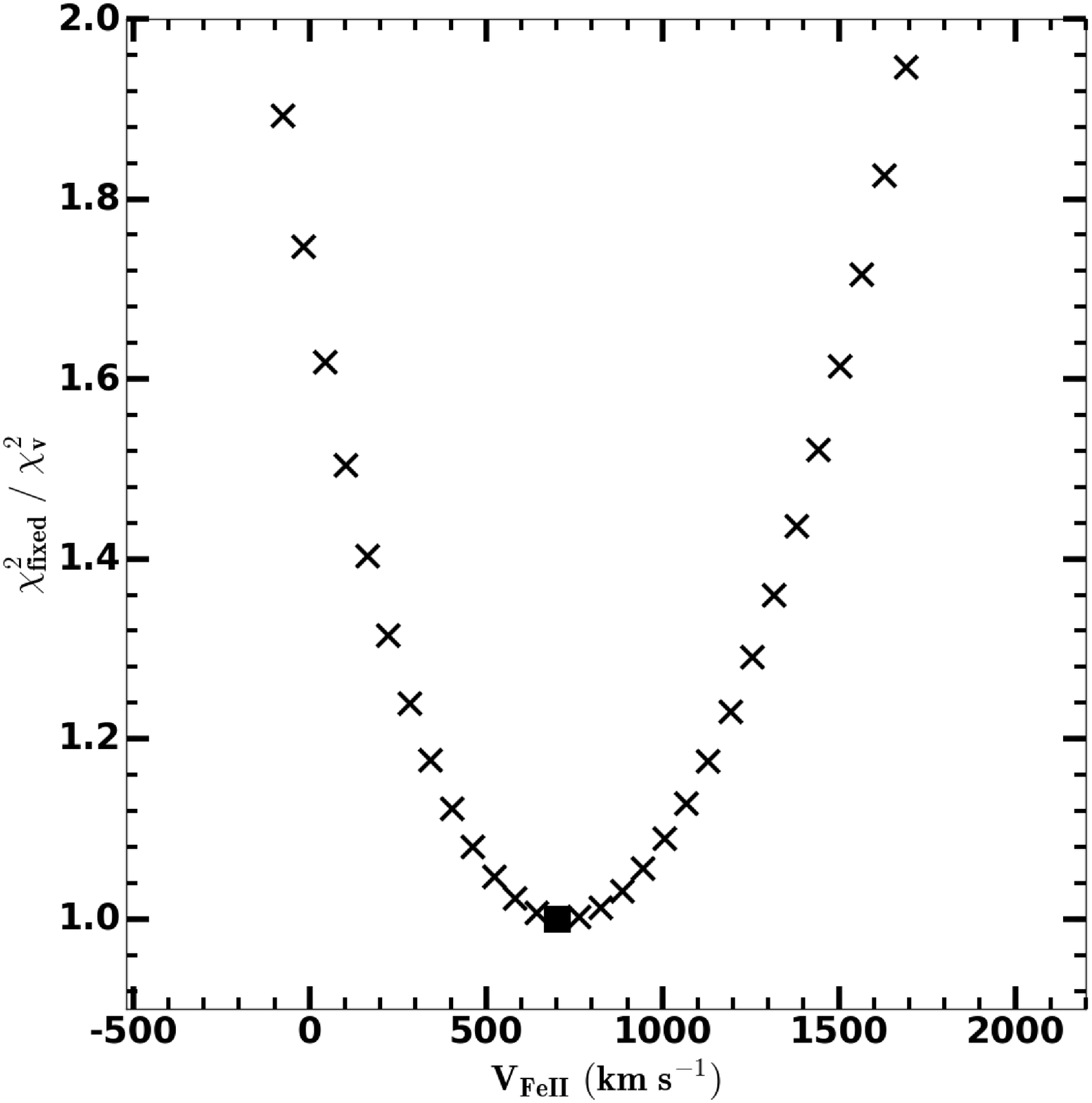}
	\includegraphics[width = 0.4\textwidth]{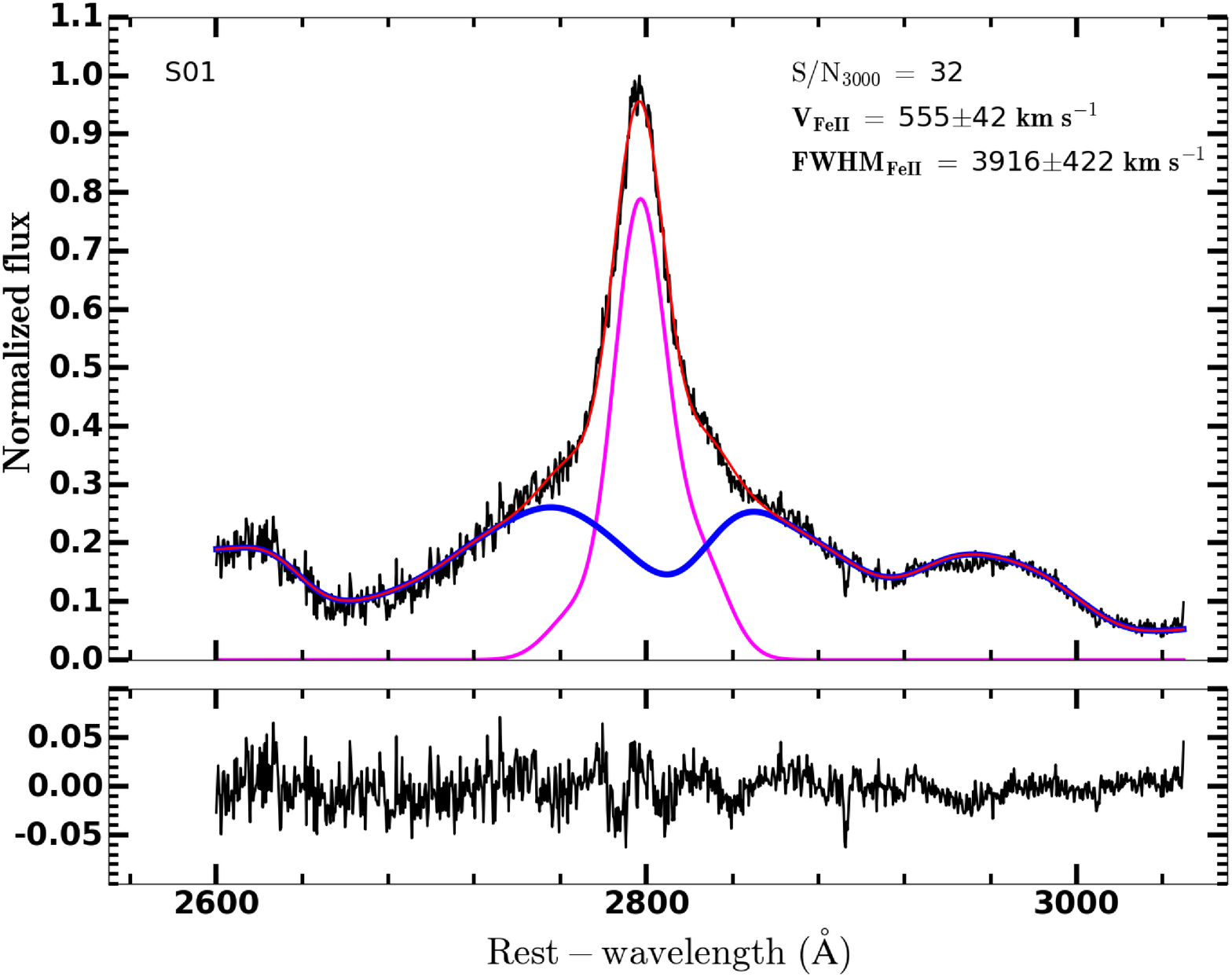}
	\includegraphics[width = 0.315\textwidth]{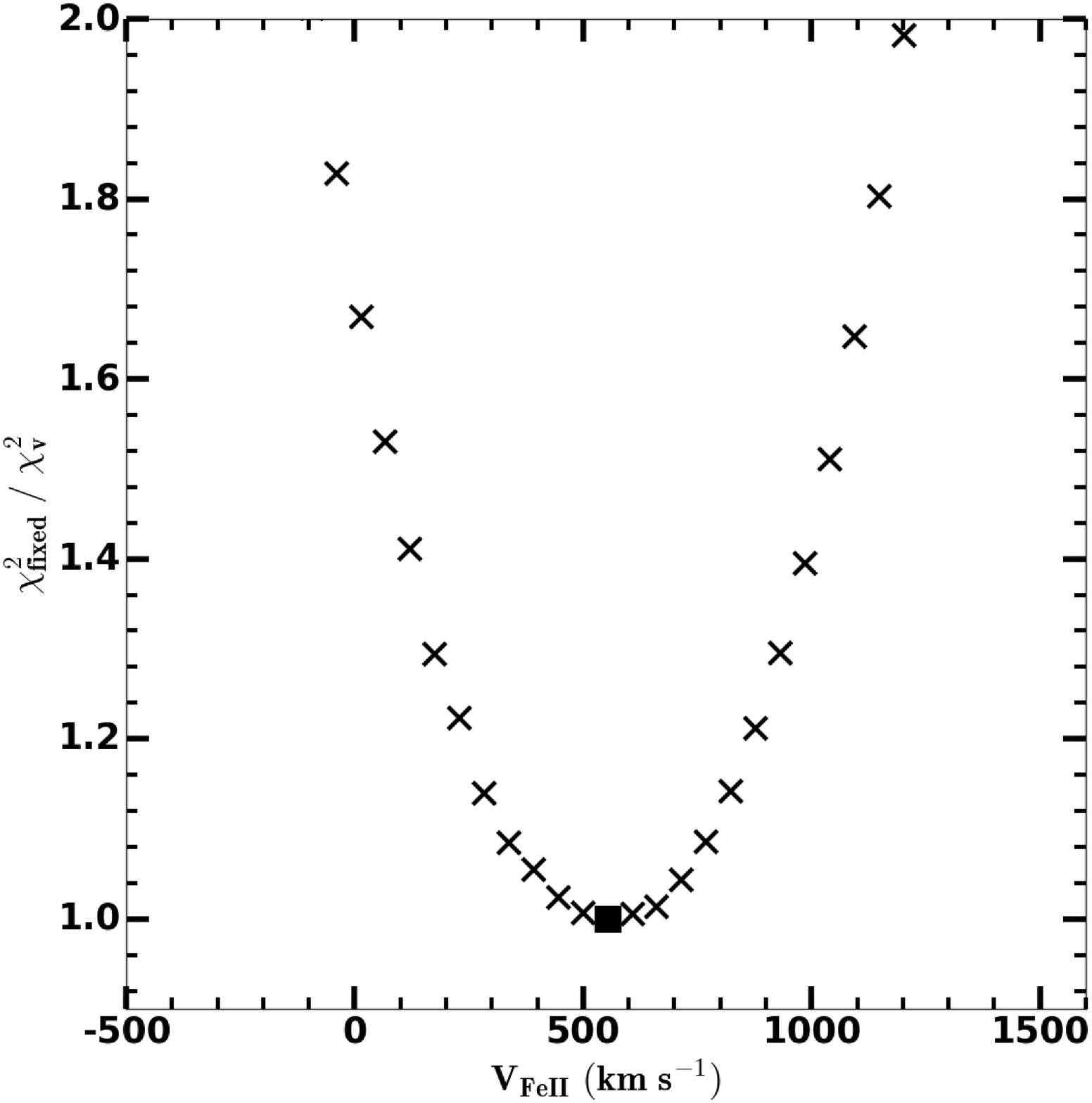}
	\caption{S01 optical and UV spectra of Keck sample. Top left panel: The rest-frame normalize continuum-subtracted spectra (thick black) and best-fit model (red), adding \FeII\ emission (blue), broad components (cyan), narrow components (magenta). In the bottom, the residual between the observed spectrum and the combined models is shown (black). Bottom left panel: similarly as color schemes in the top left panel, excepting that magenta shows the best-fit model of \MgII. Top and bottom right panels: $\chi^{2}$ curve as a function of the fixed $\rm V_{FeII}$ of the \FeII\ template. The best-fit $\rm V_{FeII}$ is shown in solid square.}
	\label{fig:spectra}
\endcenter
\end{figure*}

\section{Measurements}\label{section:meas}

We measured the kinematic properties (line width and velocity shift) and fluxes of the UV and optical \FeII\ emission lines based on the multi-component spectral analysis, following the procedure in our previous studies \citep{Woo06, McGill+08, Park+15,Woo+18}. In this section, we briefly describe the fitting process for measuring the properties of the UV and optical \FeII\ emission.

\subsection{The UV Spectra Fitting}\label{section:uv}

The fitting process of the UV \FeII\  emission lines (\FeUV) includes several steps: the observed spectra were fitted simultaneously with a combination of three pseudo-continuum components: a power-law, a Balmer continuum, and a Gaussian-velocity-convolved \FeII\ template. 
We used a simple power-law for a continuum from an accretion disk:
 \begin{equation}
	f_{\lambda } \propto \lambda ^{\alpha }
 \end{equation}
 where $\alpha$ is a power-law slope. 
 The Balmer continuum \citep{Grandi82} is calculated as:
 \begin{equation}
	f_{\lambda }^{BaC} = f_{norm}^{BE}B(\lambda ,T_{e})[1-e^{-\tau _{BE }(\frac{\lambda }{\lambda _{BE}})^{3}}]
 \end{equation}
 where $B(\lambda ,T_{e})$ is Plank blackbody spectrum at the electron temperature $\rm T_{e}$ of 15,000 K. $\rm \tau_{BE}$ is optical depth at the Balmer edge, $\rm \lambda _{BE}$ = 3646\AA. $\rm f_{norm}^{BE}$ is a normalized flux density. 

For \FeII\ emission blends, we adopted the \FeII\ template based on the I Zw 1 from \citet{Tsuzuki06}.  By convolving the \FeII\ template with a Gaussian function, we generated a series of  \FeII\ models and fitted the observed \FeII\ with the modes, including a velocity shift as a free parameter. The process was performed in the wavelength range of 2600\AA\ $-$ 3090\AA. After subtracting the pseudo-continuum from the observed spectra, we fitted the \MgII\ line by using a sixth order Gauss-Hermite series (see Section 3.2 in \citealp{Woo+18}). The best-fit models were determined by $\chi^{2}$ minimization using the nonlinear Levenberg-Marquardt least-squares fitting routine technique, MPFIT \citep{markwardt09}. 

\subsection{The Optical Spectra Fitting}\label{section:opt}

Similarly, we modeled the optical region with a combination of three components: a single power law, an \FeII\ template, and a host galaxy template. The fitting was applied in the regions of 4430\AA\ $-$ 4770\AA\ and 5080\AA\ $-$ 5450\AA. We used the I Zw 1 \FeII\ template \citep{BG92} and the stellar template from the Indo-US spectral library \citep{Valdes04}. We convolved the \FeII\ and host galaxy templates with a Gaussian and determined the velocity shifts and line width of the optical \FeII\ (\FeOPT) emission blends as well as those of stellar absorption lines. After subtracting the pseudo-continuum from the observed spectra, we fitted the broad component of the \Hb\ line by using a sixth order Gauss-Hermite series (see Section 3.1 in \citealp{Woo+18}). The best-fit models were also archived using the $\chi^{2}$ minimization with the MPFIT.

\begin{figure*}
\figurenum{2}
\center
	\includegraphics[width = 0.44\textwidth]{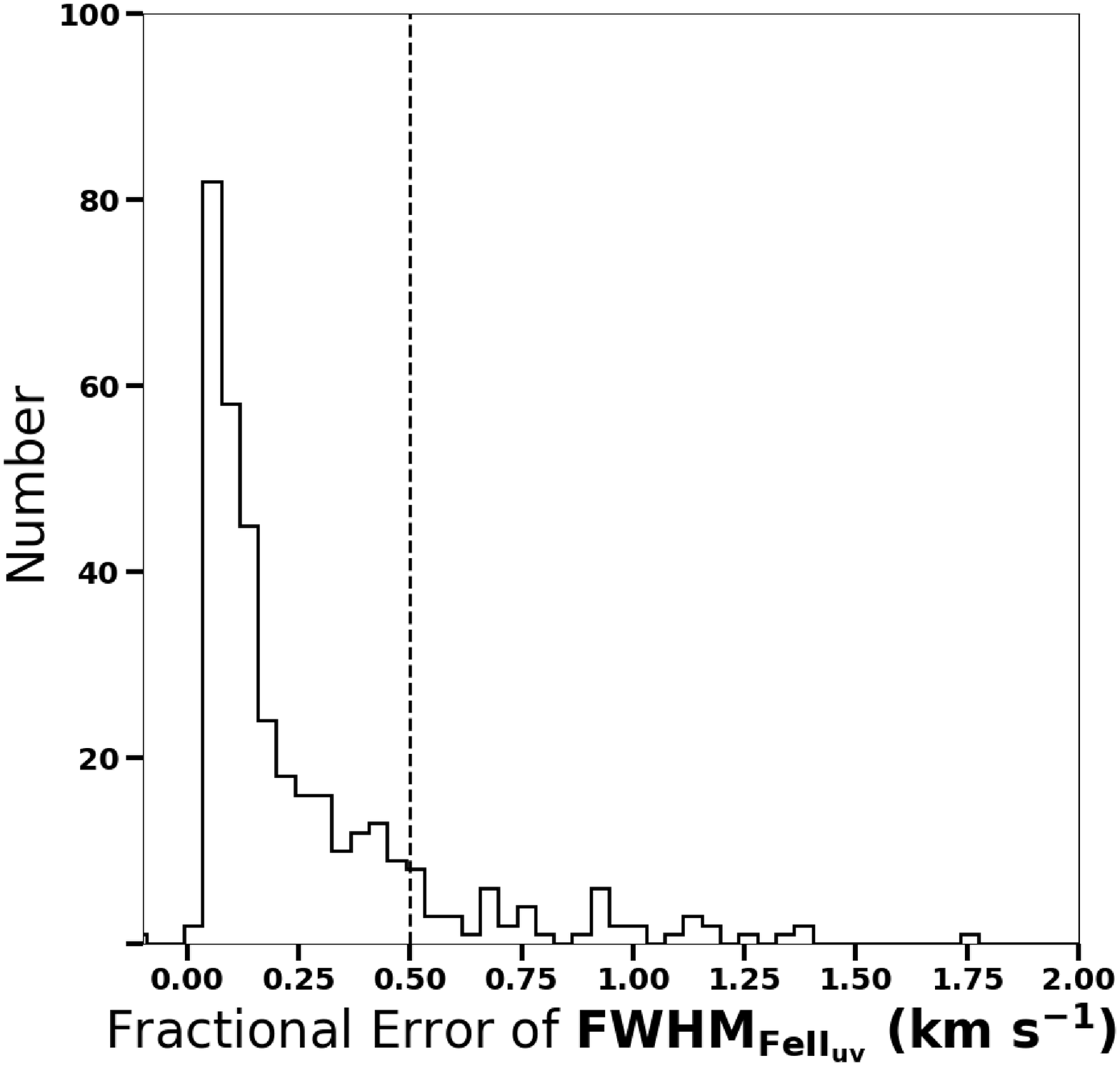}
	\includegraphics[width = 0.44\textwidth]{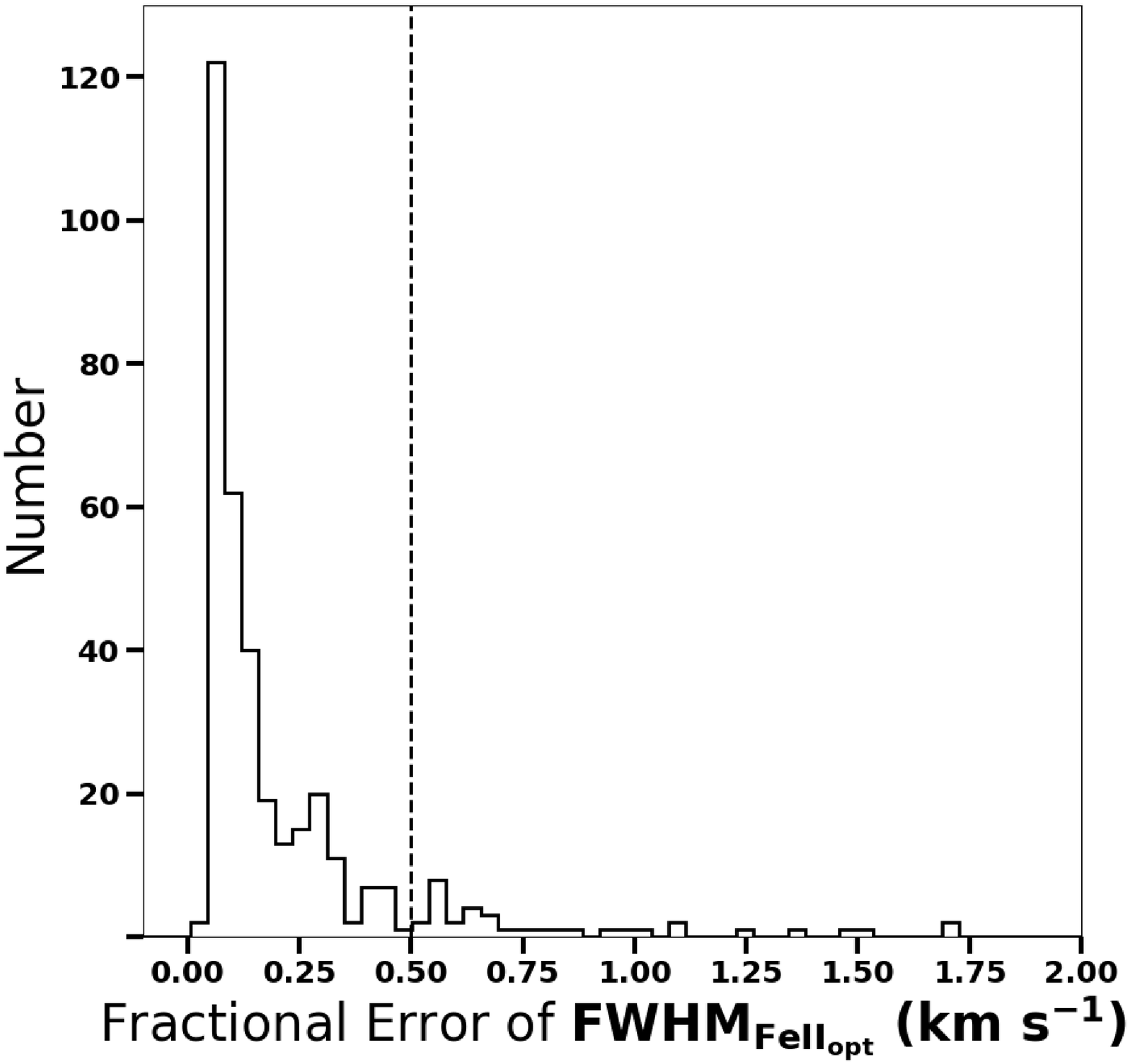}
	\includegraphics[width = 0.43\textwidth]{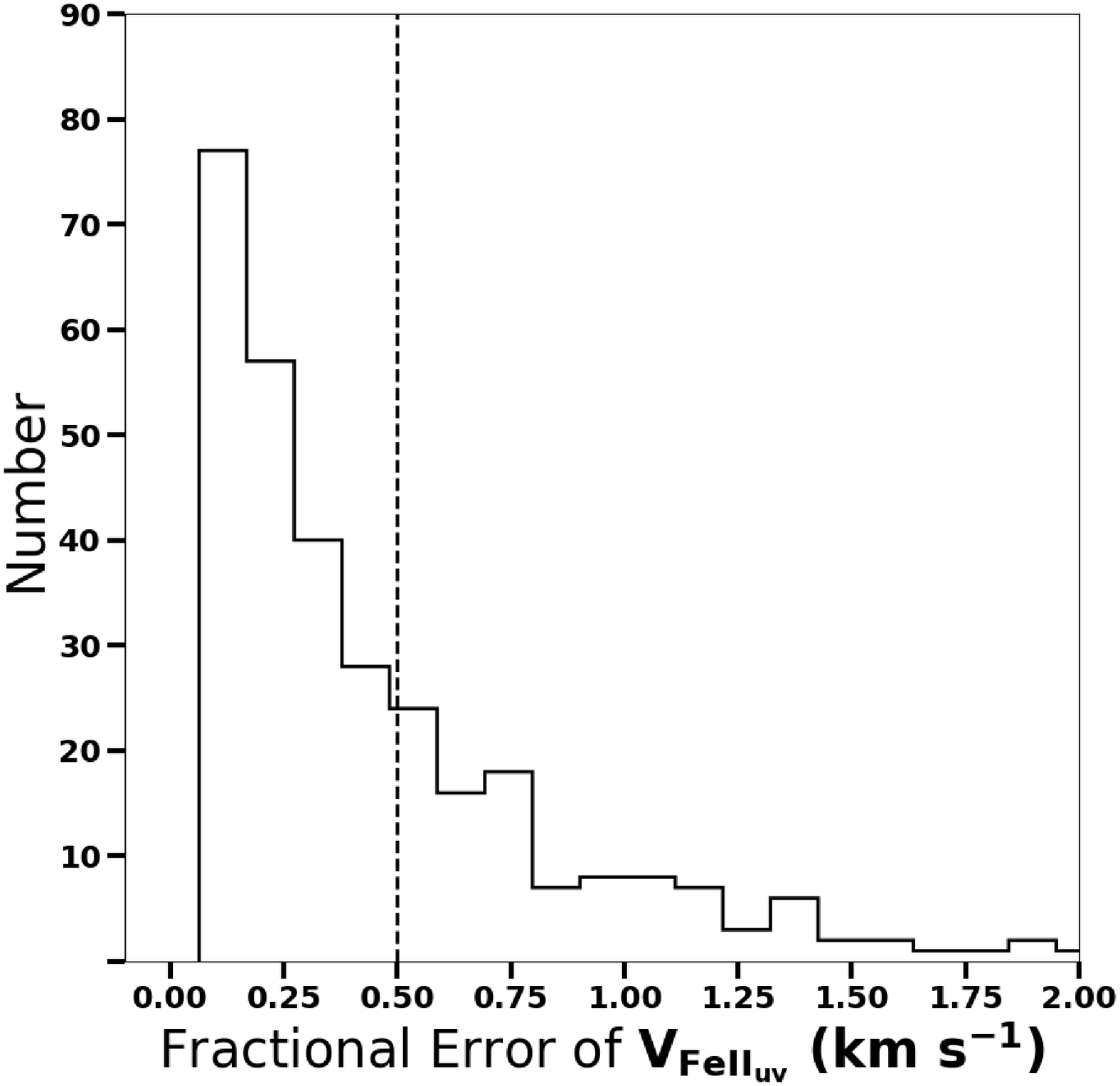}
	\includegraphics[width = 0.43\textwidth]{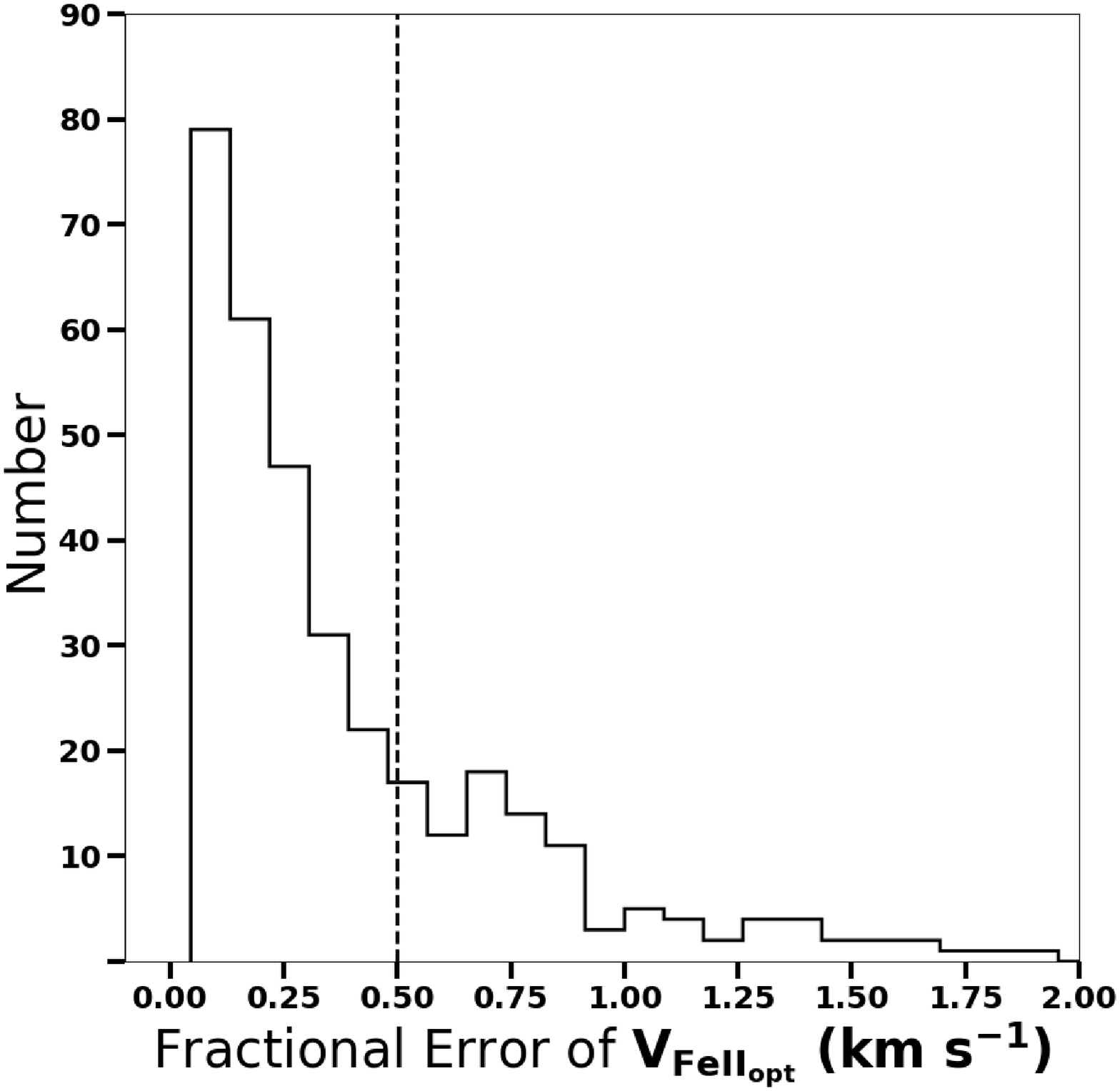}
	\caption{Fractional error distributions of FWHM and velocity of \FeUV\ and \FeOPT, respectively. The dash lines show the fraction value of 0.5.}
	\label{fig:kinematic_error}
\endcenter
\end{figure*}

\begin{figure*}
\figurenum{3}
\center
	\includegraphics[width = 0.3\textwidth]{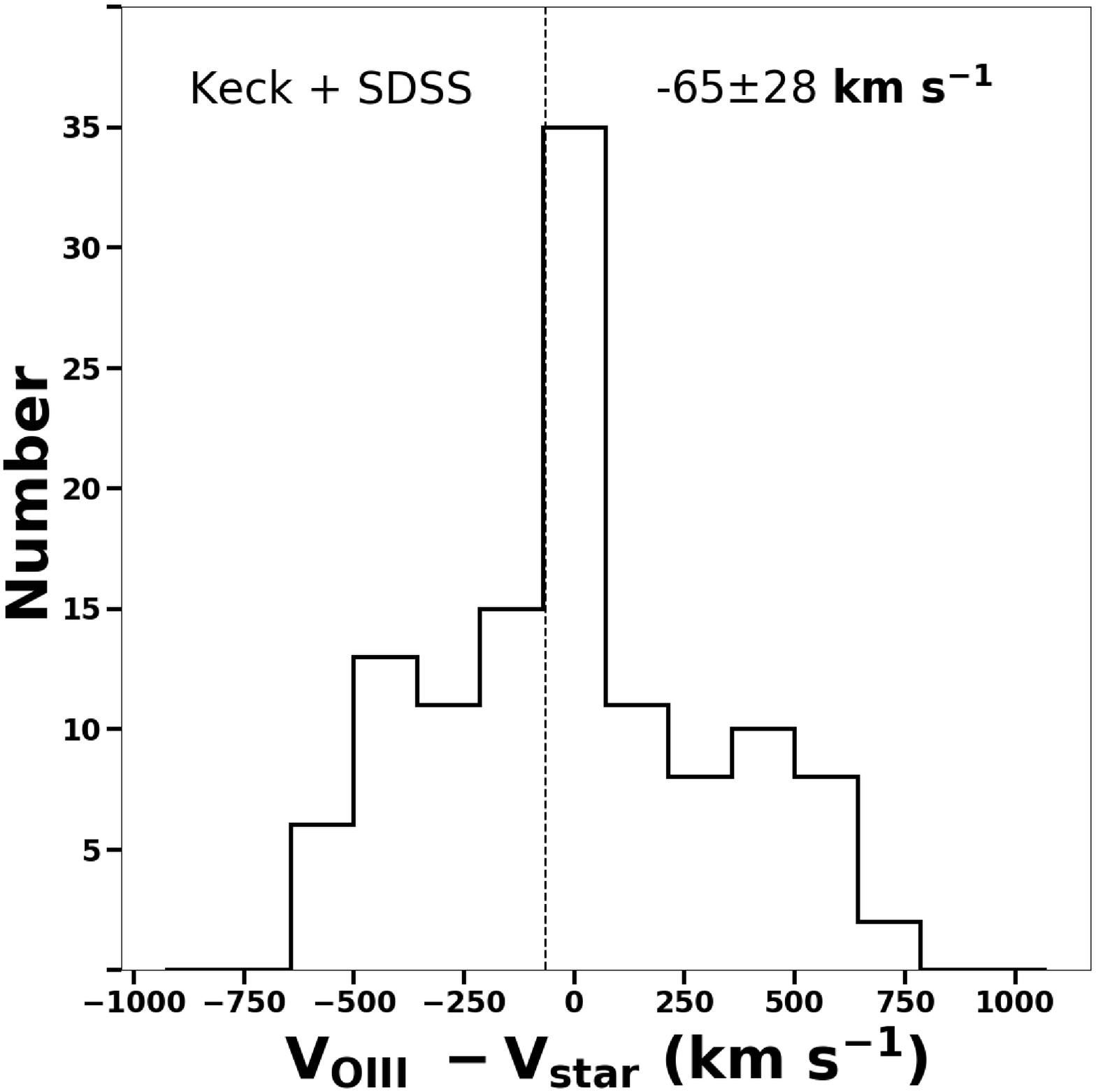}
	\includegraphics[width = 0.3\textwidth]{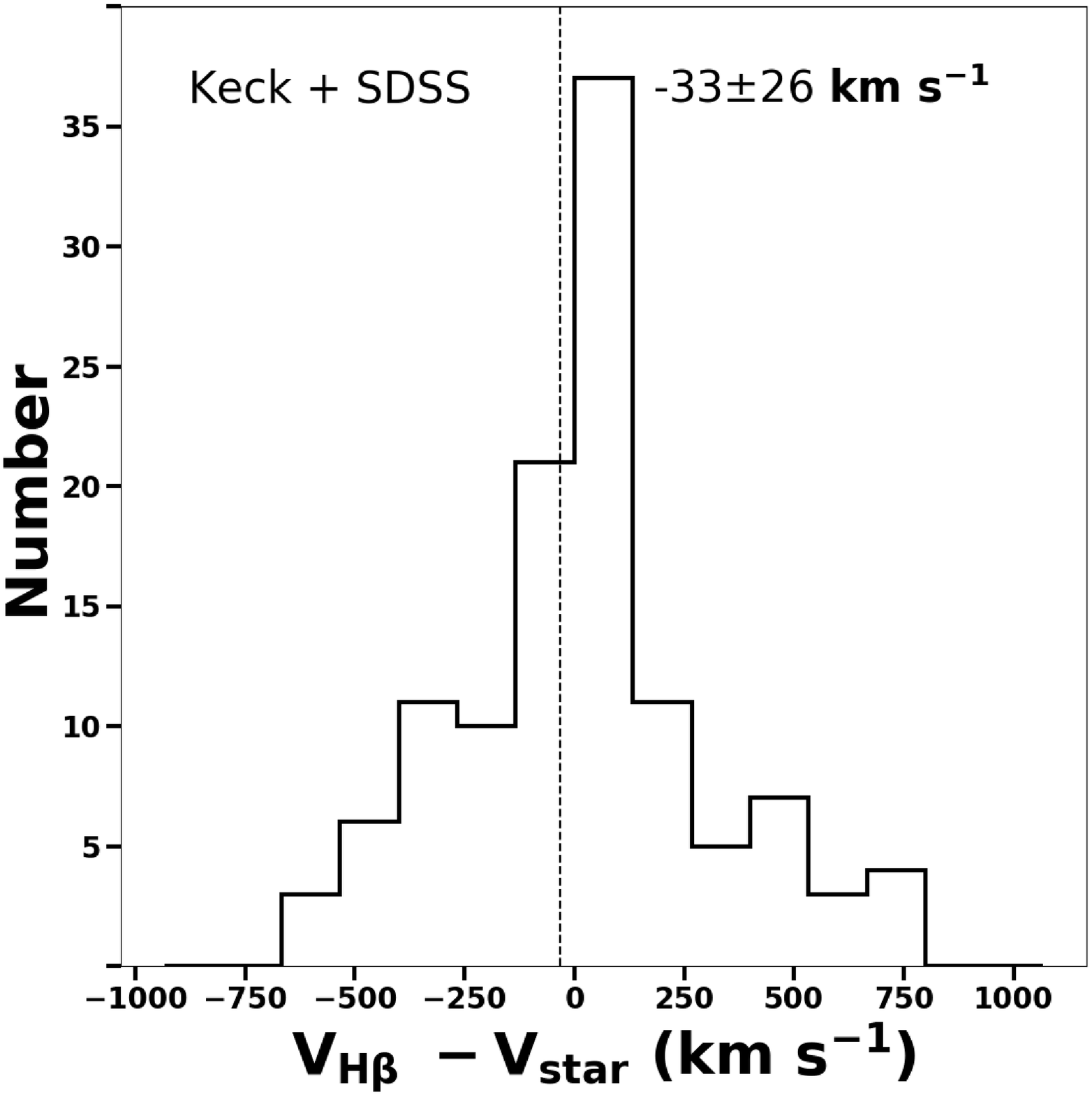}
	\includegraphics[width = 0.3\textwidth]{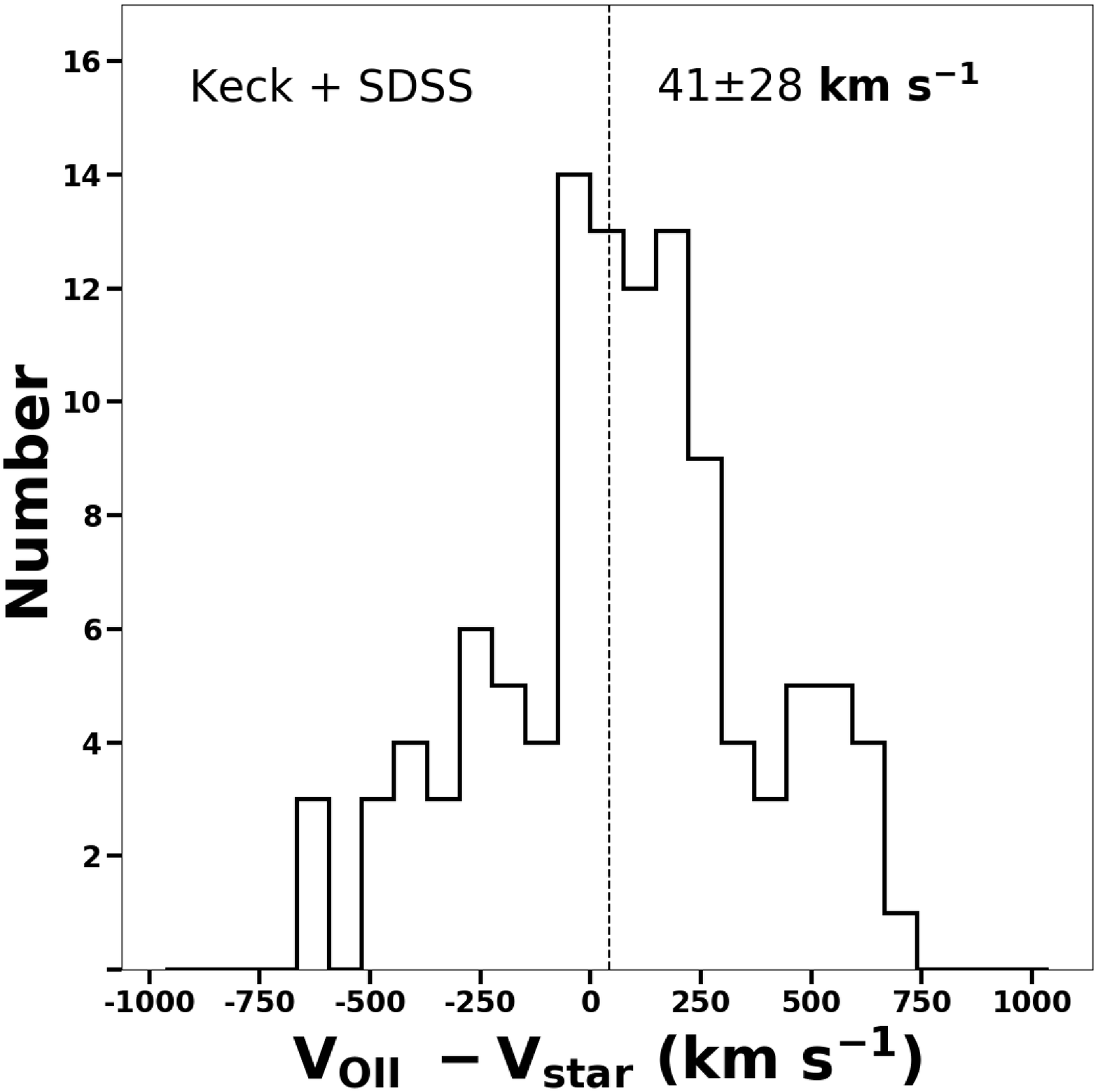}
	\caption{Distribution of the velocity shift of the peak of \OIII\ (left panel), the peak of \Hb\ (middle panel), and the peak of \OII\ (right panel), with respect to the systemic velocity measured based on stellar absorption lines, using our Keck and SDSS samples. The dash-line shows the mean of each distribution.}
	\label{fig:hist_keck}
\endcenter
\end{figure*}

\subsection{\FeII\ Emission Line Widths and Velocity Shifts}{\label{section:vel_width}}

The line width and velocity shift of the UV and optical \FeII\ emission are determined in the fitting process, where we used a series of 
FeII template with a various width and a velocity shift as free parameters.
For the Keck spectra, the measured line widths were corrected for the instrumental resolution of $\sim$145 \kms\ and $\sim$55 \kms\ in line dispersion, respectively for the UV and optical spectra, by subtracting the instrumental resolution from the measured velocity in quadrature. 
We also corrected for the instrumental resolution $R$ $\sim$ 1800 and 2200 of the SDSS spectra.
In Figure \ref{fig:spectra} we present examples of the best-fit result in the UV and optical ranges, which 
confirms the systemic redshift of \FeII\ emission lines in both UV and optical. To estimate the confidence of the fit, we used $\chi^{2}$ statistic recipe as described in Section 11.4 of \citet{Bevington03} and Appendix A of \citet{Hu+12}. We considered the fit with displacement of fixed $\rm V_{FeII}$ from -500 to 2000 \kms.

The measurement errors of line widths and velocity shifts were determined based on the Monte Carlo simulations. We generated 100 mock spectra, for which the flux at each wavelength was randomized by the flux error. Then, we applied the same fitting method for each spectrum. We adopted 1$\sigma$ dispersion of the distribution of the measurements as the error. Figure \ref{fig:kinematic_error} shows the fractional error of line widths and velocity shifts of \FeII\ emission lines.
To avoid uncertain measurements, we decided to remove the targets with the fractional error of line width or velocity shift larger than 50\%. 

\subsection{Systemic Velocity}{\label{section:v_ref}}

To measure the velocity shift of the UV and optical FeII emission blends, we first need to determine the systemic velocity of each target. While the systemic velocity can be best measured based on stellar absorption lines, luminous AGNs typically do not present strong stellar absorption lines. 
Instead, we consider the peak of the \OIII\ 5007\AA\ line for determining the systemic velocity as various previous studies have performed (e.g., \citealp{Hu+08}; \citealp{Hu+12}; \citealp{Sulentic+12}; \citealp{Kovacevic+15}). 
It is well known that the \OIII\ line manifests outflows in type 1 and 2 AGNs \citep[e.g.,][]{Bae&Woo14, Woo+16, Woo+17, Rakshit+18}. 
By using flux weighted center (first moment) of \OIII, various previous studies determined the velocity shift of \OIII\ with respect to stellar absorption lines. However, the peak of \OIII\ does not show a large velocity shift and can be used as a proxy for stellar absorption lines albeit with large uncertainty. 
By selecting AGNs, which show a relatively strong stellar component in the continuum ($>$ 30\% in the total flux),
we tested the difference between the systemic velocity respectively measured based on stellar absorption lines and the peak of \OIII\ in Figure \ref{fig:hist_keck} (left panel). 
The difference is relatively small with an average of -65 $\pm$ 28 \kms, while the worst case shows several hundred \kms, suggesting that we may use the peak of the \OIII\ line for determining systemic velocity since it is close to the stellar velocity.

We also consider the peak of the \Hb\ line for determining the systemic velocity of each target. Using AGNs with strong stellar absorption lines, we compared the stellar-line-based systemic velocity with the peak of \Hb\ (Figure \ref{fig:hist_keck}, middle panel). The difference is an average of -33 $\pm$ 26 \kms, which is smaller than the case of the peak of \OIII, presumably due to the effect of outflows on the \OIII\ line profile. 
Note that careful attention is required in using the peak of \Hb\ to infer the systemic velocity, since for individual AGNs the peak of \Hb\ may not be
a good tracer of systemic velocity.

In the case of the UV spectra of Keck data, we cannot use \Hb\ since the UV spectra were obtained independently with a different spectrograph, hence, there could be a systematic shift between UV and optical spectra. Thus, we used the peak of the \OII\ 3727\AA\ line to determine the systemic velocity. 
Figure \ref{fig:hist_keck} (right panel) compares the systemic velocity measured from stellar absorption lines and from the peak of \OII. 
For most targets, the difference is relatively small with an average of $\sim41 \pm 28$ \kms. As in the case of using the \Hb\ peak, the difference between the peak of \OII\ and stellar velocity can be large up to $\pm$750 \kms\ for individual AGNs.
In summary, we used the peak of \Hb\ for measuring the systemic velocity ($\rm V_{ref}$) in our analysis. In the case of the 38 objects with the Keck UV spectra, we used the peak of \OII\ for $\rm V_{ref}$. There are two targets with the Keck spectra, which show large difference in the systemic velocity measured from \Hb\ and \OII\ ($>-500$ \kms, SS12 and SS17). We removed these two targets when we compare the velocity shift of \FeUV\ emission lines to other physical quantities.

\subsection{\FeII\ Emission Line fluxes}

In addition to the kinematical properties, we also measure the line fluxes based on the \FeII\ emission lines. Following previous studies in the literature, the line flux of the optical \FeII\ is integrated over a spectral range of 4434-4684 \AA\ using the best-fit model (see \citealp{Woo15}). In the case of the UV \FeII, the line flux is calculated by summing over a spectral range of 2600-3050 \AA\ as similarly done by \citet{Kovacevic+15}.

\section{Results}\label{section:result}

\subsection{Comparing UV and Optical \FeII\ Emission}

In this section, we investigate the kinematical properties manifested in the UV and optical \FeII\ emission lines to understand the connection between two emission line regions.

\begin{figure}
\figurenum{4}
\center
	\includegraphics[width = 0.45\textwidth]{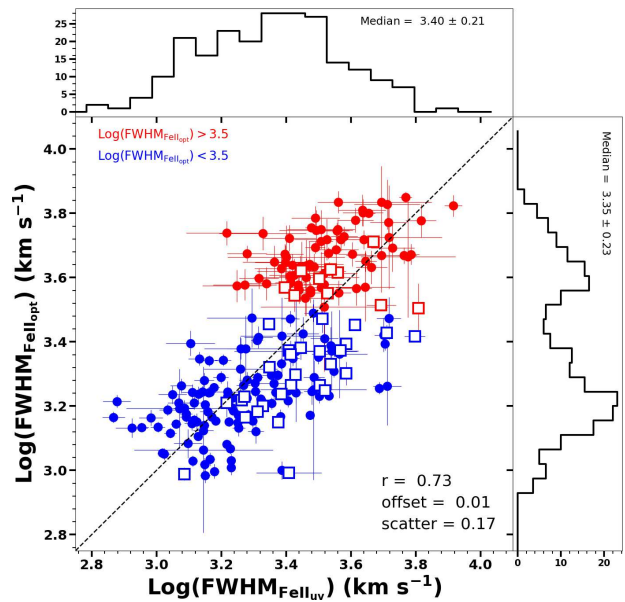}
	\caption{Comparison of FWHM of \FeUV\ and \FeOPT. Squares indicate Keck sample with S/N > 20 in the continuum at 3000 \AA\ and 51000 \AA\ . Filled circles show the measurements from SDSS sample. Black dash-lines denote $\rm FWHM_{FeII_{opt}} = FWHM_{FeII_{uv}}$. The correlated coefficient, offset and rms scatter from the black dash line are shown in the plots. Blue samples are $\rm FWHM_{FeII_{opt}} < 10^{3.5}$ (Group A), and red samples are $\rm FWHM_{FeII_{opt}} > 10^{3.5} $ (Group B).}
	\label{fig:fwhm_uvopt}
\endcenter
\end{figure}

\subsubsection{Line Widths of \FeII\ Emission Lines}

We compare the line width measured from the UV and optical \FeII\ emission lines in Figure \ref{fig:fwhm_uvopt}. The FWHM of UV \FeII\ ranges from $\sim$1000 to $\sim$10, 000 \kms, with the mean log FWHM (\kms) = 3.40 $\pm$ 0.21. In the case of \FeOPT, the mean log FWHM (\kms) is 3.35 $\pm$ 0.23. The comparison shows $\sim$0.2 dex scatter and similar line widths between
UV and optical lines. To explore the correlation of the two line widths, we perform the Spearman's rank-order correlation, 
finding that \FeOPT\ and \FeUV\ lines show are correlated with a correlation coefficient r = 0.73. 
The results indicate that the UV and optical \FeII\ emission lines are originated from regions which are closely located to
each other as previously reported in the literature \citep{Kovacevic+15}. 

Nevertheless, we find an interesting trend that the distribution of \FeOPT\ FWHM is divided at $\sim$3500 \kms. If 
we separate the sample into two groups: Group A: FWHM of \FeOPT\ $<$ $\sim$3200 \kms, and Group B: FWHM of \FeOPT\ $>$ $\sim$3200 \kms, then Group A and Group B show somewhat different comparison. While the line width of \FeII\ is comparable between UV and optical in Group A, \FeOPT\ is broader than \FeUV\ by $\sim$0.1 dex in Group B. Thus, we compare the properties of UV and optical \FeII\ using these two separate groups in the following sections.

\begin{figure}
\figurenum{5}
\center
	\includegraphics[width = 0.45\textwidth]{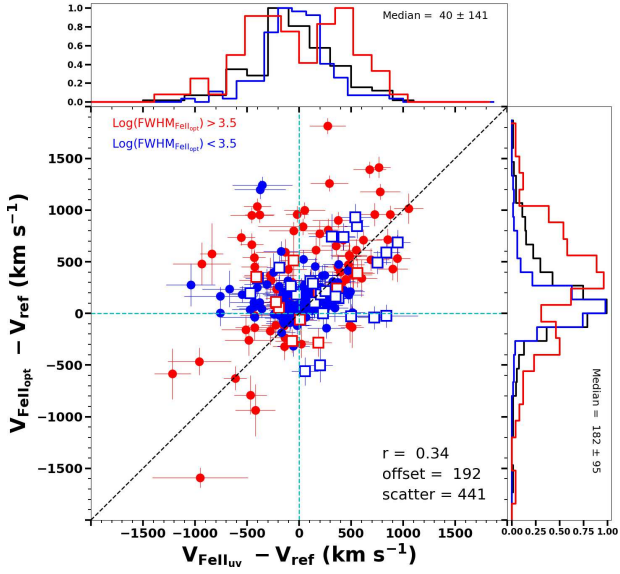}
	\caption{Comparison of velocity shifts of \FeUV\ and \FeOPT. Distribution velocity shifts of all sample (black color), Group A (blue color) and Group B (red color) are shown in the histogram panels. Squares indicate Keck sample with S/N > 20 in the continuum at 3000 \AA\ and 5100 \AA. Filled circles show the measurements from SDSS sample. Cyan dash-lines show the zero velocity shift. Black dash-lines denote $\rm V_{FeII_{opt}} = V_{FeII_{uv}}$. The correlated coefficient, offset and rms scatter from the black dash line are shown in each panel.}
	\label{fig:kinematic_uvopt}
\endcenter
\end{figure}

\subsubsection{Velocity Shifts of \FeII\ Emission Lines}\label{section:confirm}

We compare the velocity shifts of \FeUV\ and \FeOPT\ in Figure \ref{fig:kinematic_uvopt}. For both \FeUV\ and \FeOPT, the velocity shift ranges around $\pm1500$ \kms, while the mean is 40 $\pm$ 141 \kms and 182 $\pm$ 95 \kms, respectively for \FeUV\ and \FeOPT. As described in Section \ref{section:v_ref}, we \Hb\ (and \OII) shows on velocity shift with respect to stellar absorption lines, with an average of $-33\pm26$ \kms\ and $41\pm28$ \kms, respectively. For the uncertainty of the systemic velocity, we added the 3 $\sigma$ uncertainty of the systemic velocity (i.e., a factor of three of the rms dispersion in the distribution of the \Hb\ or \OII\ velocity shift, see Figure 3) to the measurement error determined from the Monte Carlo simulations in quadrature.

The velocity shift correlates between \FeUV\ and \FeOPT, while the correlation is considerably weaker than the case of the FWHM of \FeII, 
with r = 0.34. When we examine subsamples, the velocity shifts of \FeUV\ and \FeOPT\ in Group A are comparable within the scatter. 
In the case of Group B, \FeOPT\ is on average more redshifted ($\rm V_{FeII_{opt}}$ = 426 $\pm$ 123 \kms) than \FeUV\ ($\rm V_{FeII_{UV}}$ = 22 $\pm$ 148 \kms).

Overall, there is a large range of velocity shift in both \FeUV\ and \FeOPT. The majority of AGNs show redshifted \FeOPT\ while \FeUV\ is
blue-shifted for a significant fraction of AGNs. Note that \FeUV\ and \FeOPT\ show a consistent velocity shift, indicating either inflow
or outflow motion of gas, with higher velocity of \FeOPT\ in general. There are also a small fraction of AGNs, particularly in Group B, that show an opposite sign of the velocity shift between \FeUV\ and \FeOPT, implying that \FeUV\ gas is outflowing while \FeOPT\ gas is inflowing.
Note that the measured velocity shift does not entirely depend on the average gas kinematics. The observed emission line reflects
the potential anisotropic emission, which may be caused more strongly in UV \FeII\ from clouds with higher column density (for detailed discussion, see Section 5.2).

\begin{figure*}
\figurenum{6}
\center
	\includegraphics[width = 0.45\textwidth]{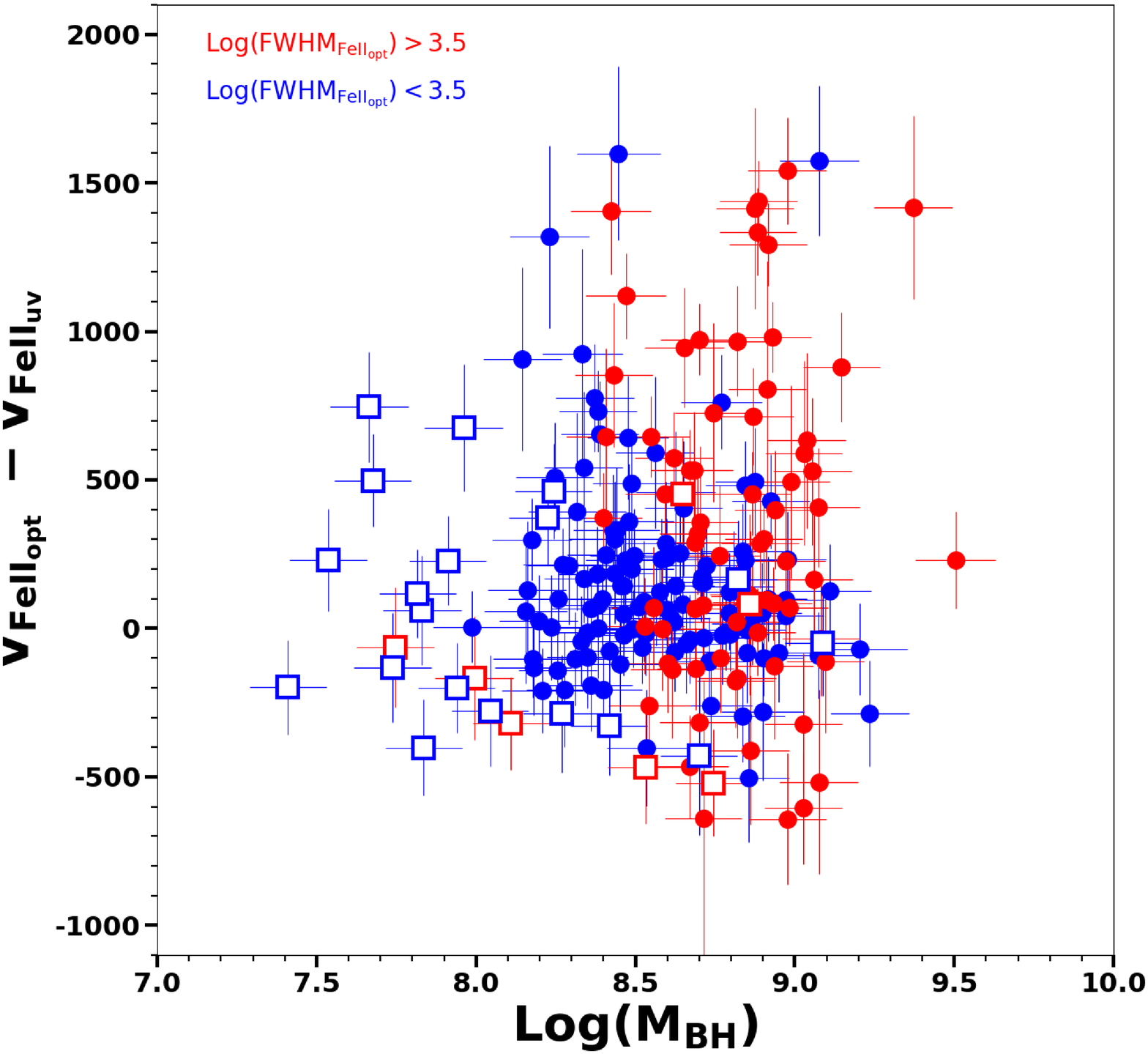}
	\includegraphics[width = 0.45\textwidth]{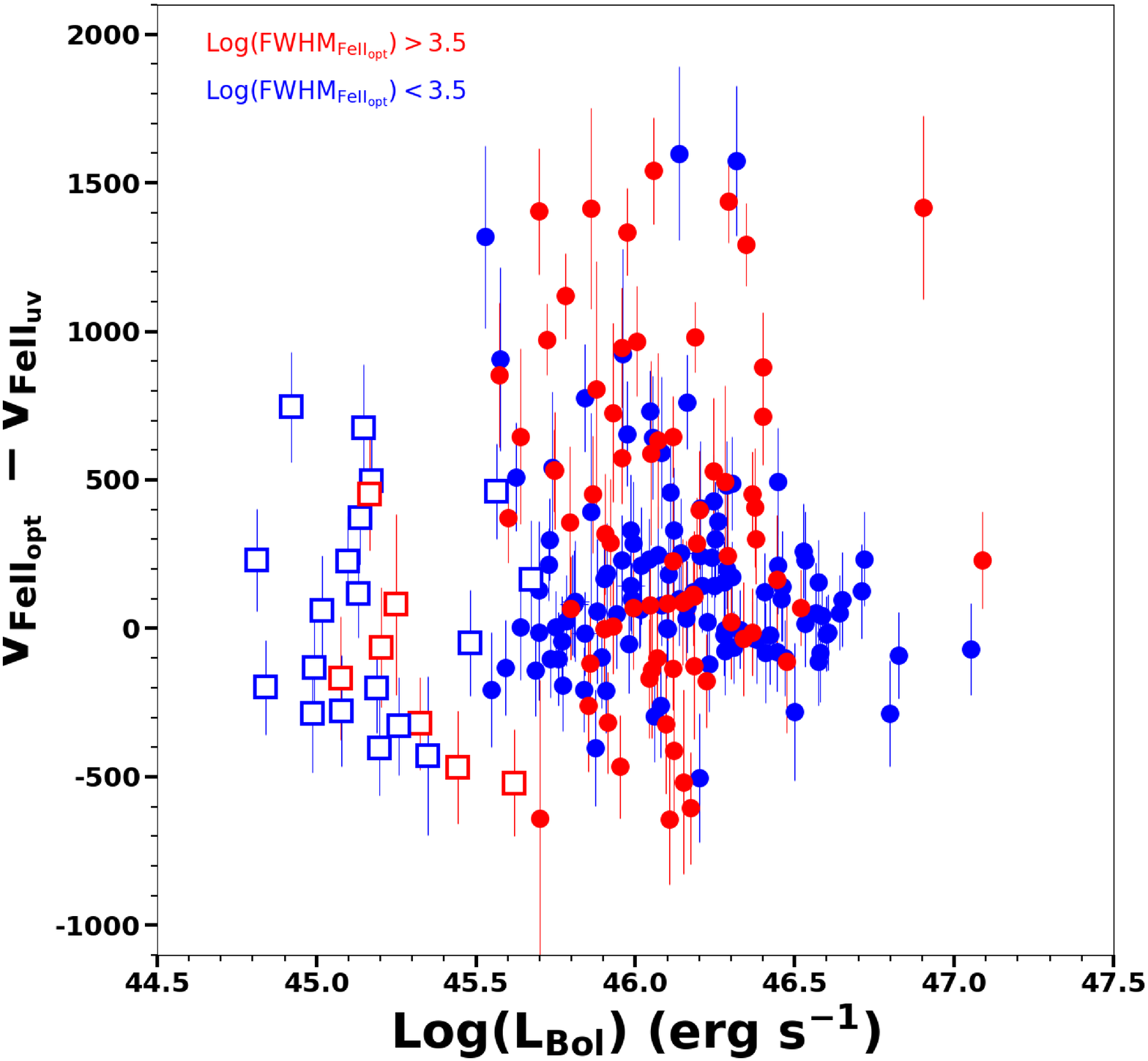}
	\includegraphics[width = 0.45\textwidth]{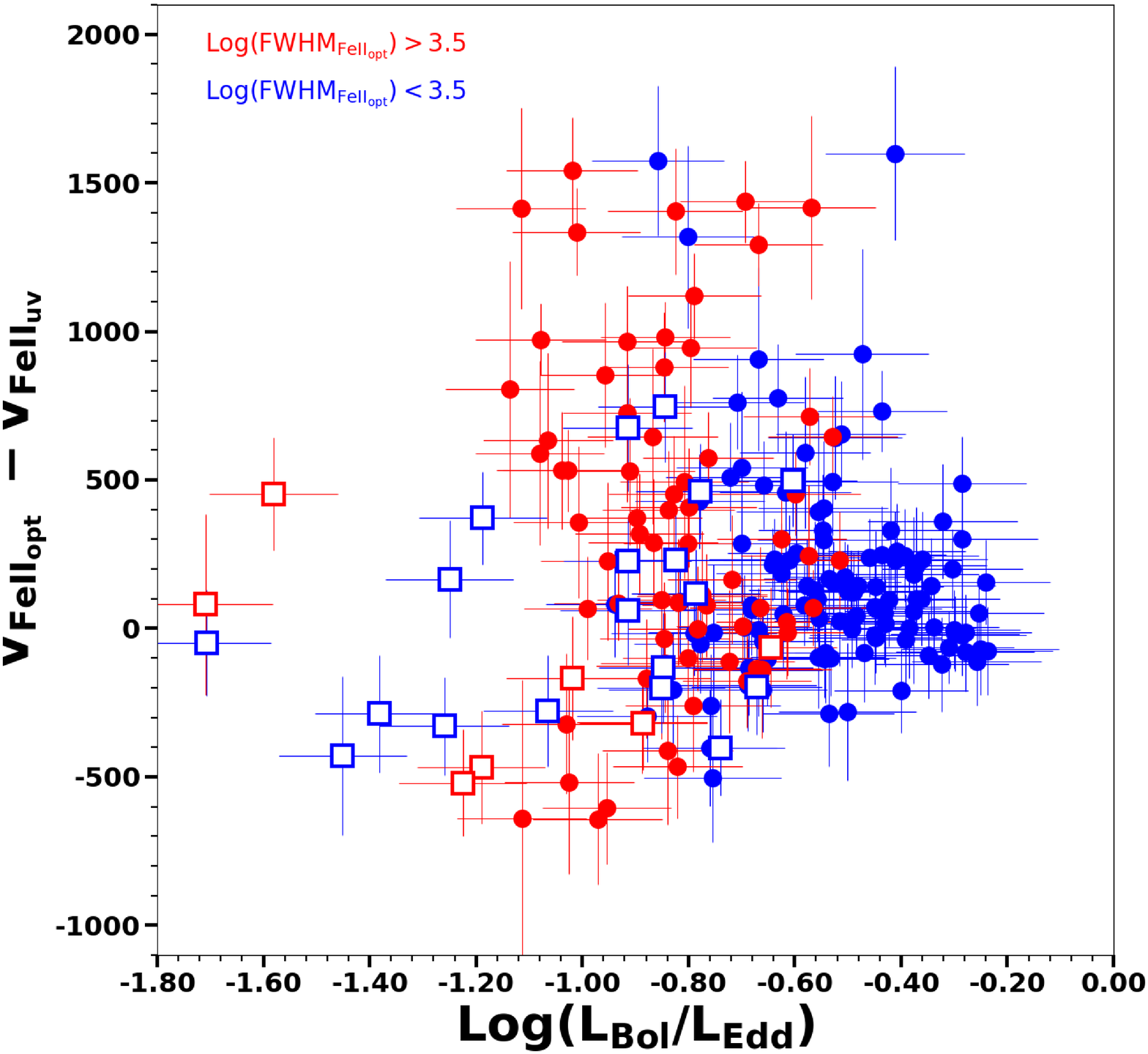}
	\includegraphics[width = 0.45\textwidth]{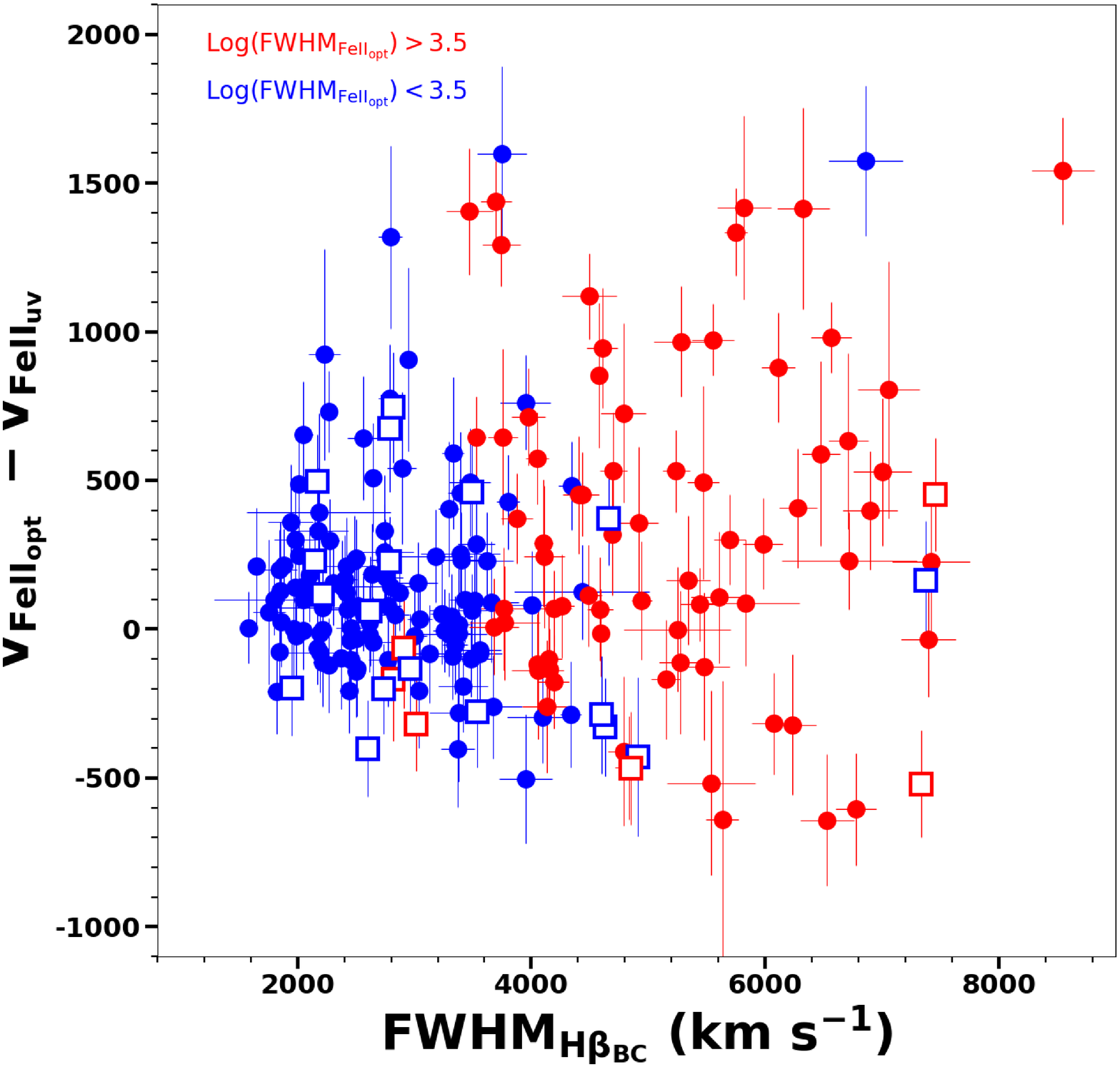}
	\caption{Comparison of the UV and optical \FeII\ velocity shift as a function of AGNs properties. Group A ($\rm FWHM_{FeII_{opt}} < 10^{3.5}$) is denoted with blue symbols, while Group B ($\rm FWHM_{FeII_{opt}} > 10^{3.5}$) is dented with red symbols. Keck and SDSS objects are presented 
	with open and filled symbols, respectively.}
	\label{fig:velocity_compare}
\endcenter
\end{figure*}



\subsubsection{\FeII\ Kinematics vs. AGNs Properties}\label{section:confirm}

To explore what is the main driver of the difference in velocity shifts between \FeUV\ and \FeOPT, we compare them with AGNs properties. In Figure \ref{fig:velocity_compare}, we show the difference in \FeUV\ and \FeOPT\ velocity shifts as a function of \mbh, bolometric luminosity, Eddington ratio, and the FWHM of the broad \Hb\ line. We find no specific trend with \mbh\ or bolometric luminosity. However, it is interesting to see the difference in the velocity shifts of Group A and B as a function of Eddington ratio luminosity, and FWHM of the \Hb\ line. Group A sample falls in the group of narrow FWHM \Hb\ (FWHM $<$ $\sim$4000 \kms), and shows relatively small difference in velocity shifts between \FeUV\ and \FeOPT, while Group B AGNs show larger FWHM \Hb\ ($>$ $\sim$4000 \kms), and display large divergence between \FeUV\ and \FeOPT\ velocity shifts. Similarly, the difference in velocity shifts becomes smaller when Eddington ratio is higher ($\rm L_{Bol}/L_{Edd}$ $>$ $\sim$-0.7), while there is a larger discrepancy for lower Eddington ratio AGNs ($\rm L_{Bol}/L_{Edd}$ $<$ $\sim$-0.7).

\subsubsection{\FeII\ vs. Other Emission lines}\label{section:compare}

\begin{figure*}
\figurenum{7}
\center
	\includegraphics[width = 0.45\textwidth]{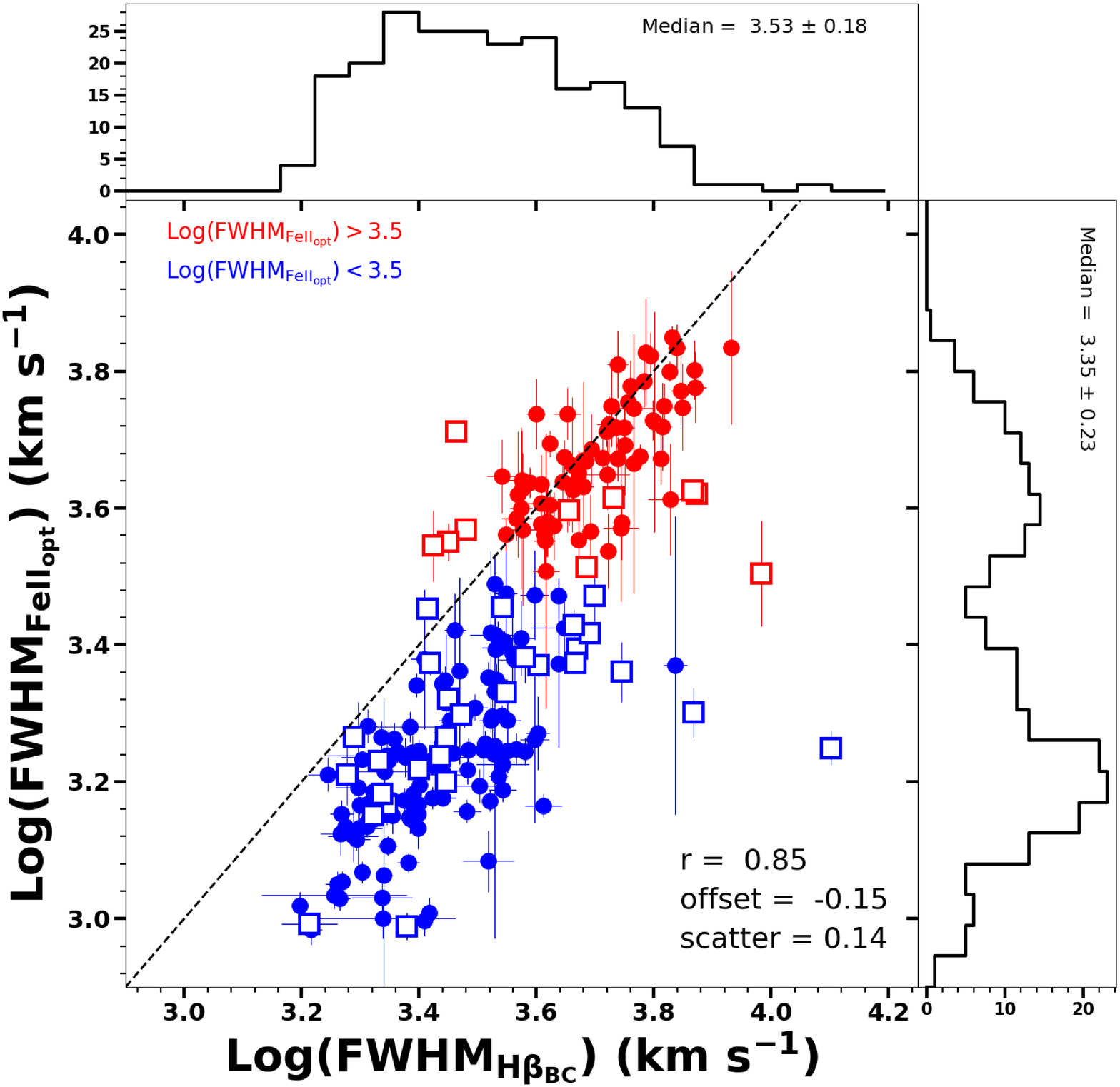}
	\includegraphics[width = 0.45\textwidth]{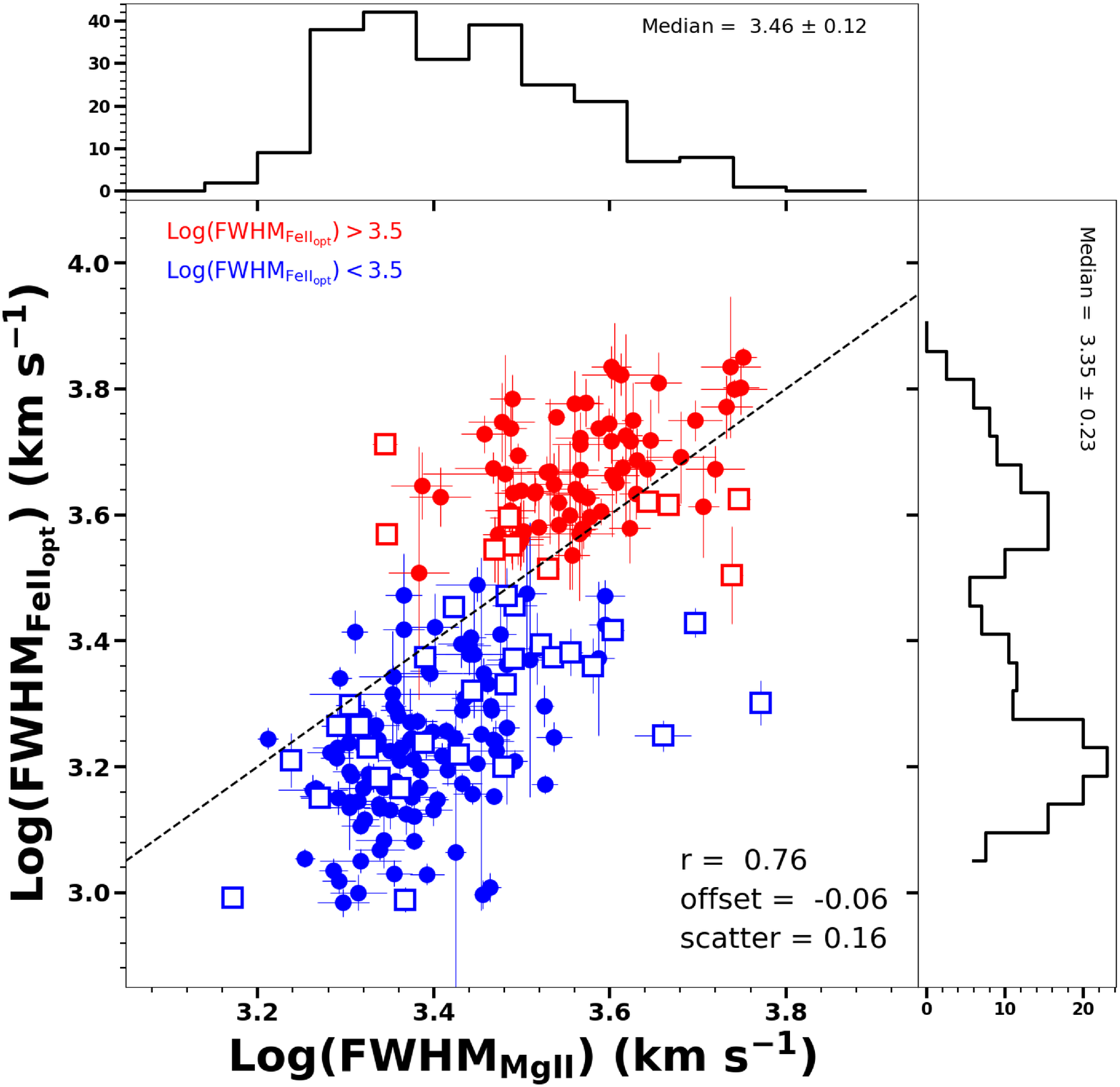}
	\includegraphics[width = 0.45\textwidth]{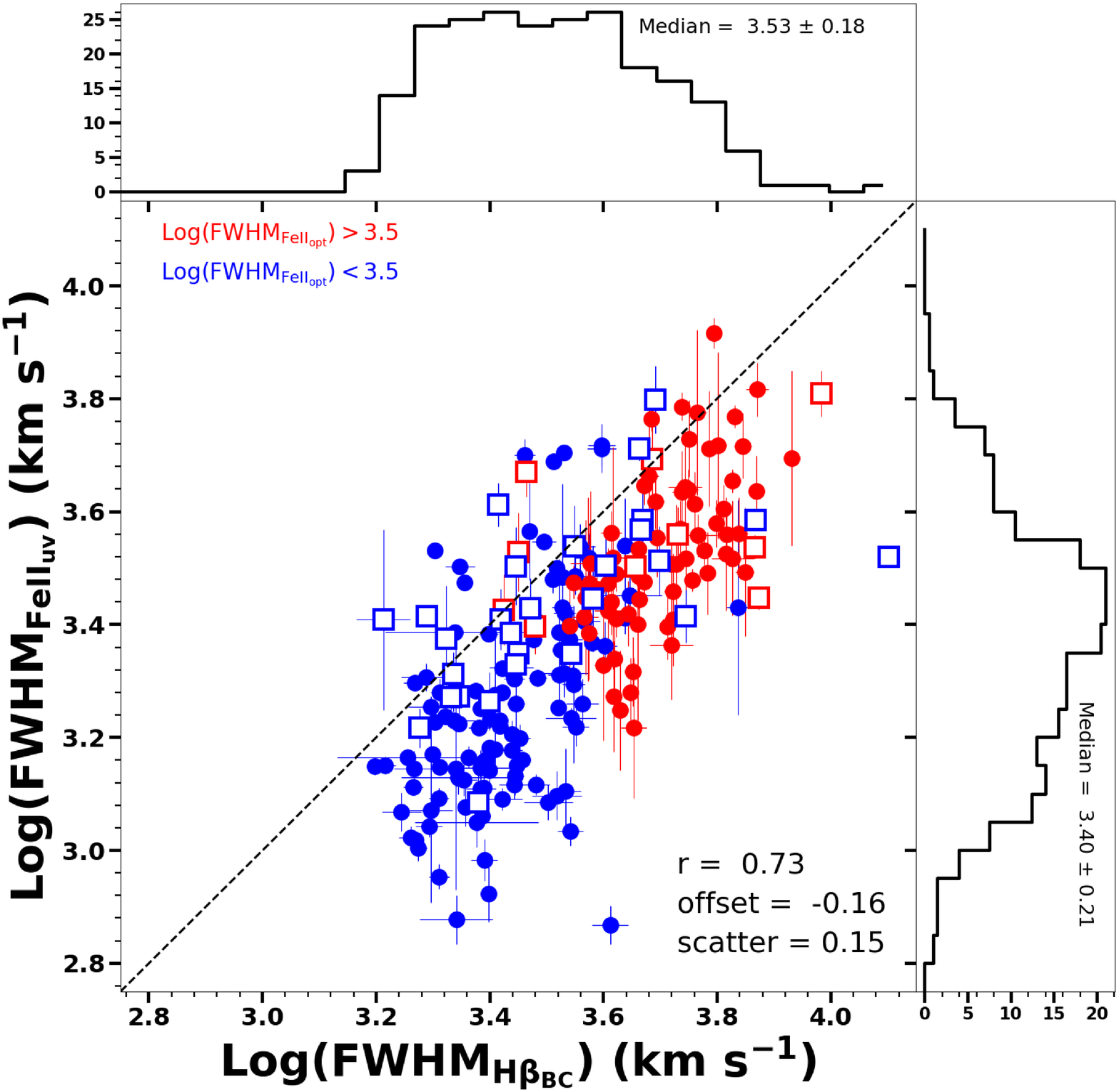}
	\includegraphics[width = 0.45\textwidth]{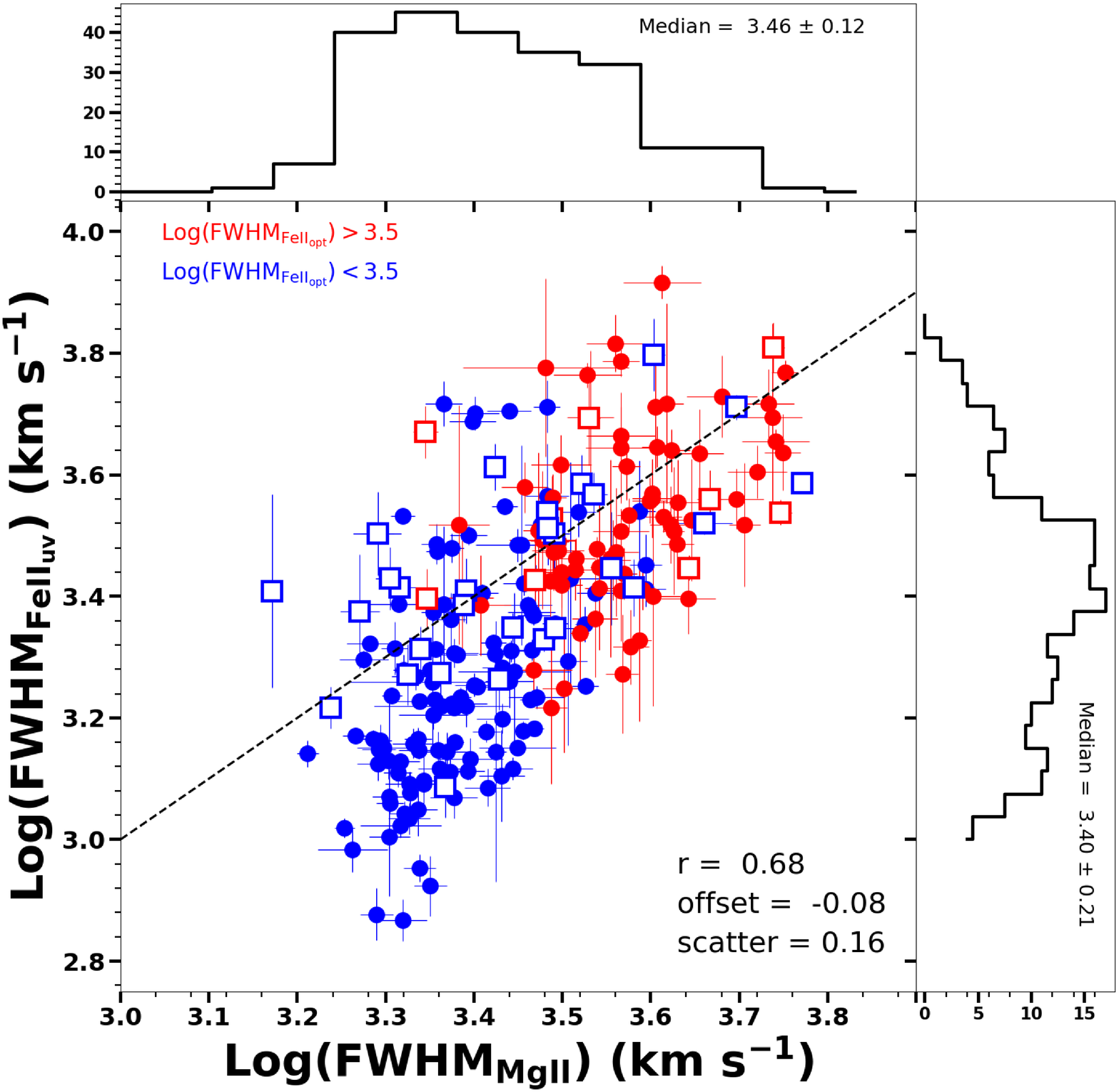}
	\caption{Comparison of FWHM of \FeUV, \FeOPT\ with that of \MgII\ and \Hb. Dash-lines denote that FWHM of the quantities of horizon and vertical axes in each panel are same. The correlated coefficient, offset and rms scatter from the black dash line are shown in the plots. Group A ($\rm FWHM_{FeII_{opt}} < 10^{3.5}$) is denoted with blue symbols, while Group B ($\rm FWHM_{FeII_{opt}} > 10^{3.5}$) is dented with red symbols. Keck and SDSS objects are presented with open and filled symbols, respectively.}
	\label{fig:fwhm_compare}
\endcenter
\end{figure*}

We compare the line width of \FeUV\ and \FeOPT\ with that of the broad emission lines, (i.e., \Hb\ and \MgII) in Figure \ref{fig:fwhm_compare}.
First, we investigate the correlation between the line width (FWHM) of \FeOPT\ and that of the broad \Hb\ line. Using the total sample, we determine
the correlation coefficient r = 0.85, indicating a good correlation between \FeOPT\ and \Hb\ line widths. We obtain the best-fit slope of 1.15 $\pm$ 0.06, which is close to a linear relationship, while we find a systemic offset of 0.15 dex, indicating that \FeOPT\ is an average narrower than \Hb\ by 30\%. 
For Group A, we find a larger systemic offset of 0.21 dex ($\sim$40\%) between \FeOPT\ and \Hb\ in Group A, implying that on average the \FeOPT\ emission region is located further out in the BLR, compared to the \Hb\ emission region. This result is consistent with the result of \citet{Hu+08a}, who found that FWHM of \FeOPT\ = $\frac{3}{4}$ FWHM of \Hb. 
In contrast, Group B shows comparable line widths between \FeOPT\ and \Hb, implying that \FeOPT\ and \Hb\ is emitted from the regions, which are close to each other.    

Second, we compare \MgII\ and \FeOPT, finding similar trends with those of \Hb. The widths of \FeOPT\ and \MgII\ are well correlated with r = 0.76. 
Again, we see the discrepancy between Group A and Group B. For Group A, \FeOPT\ is on average narrower than \MgII\ by 0.15 dex ($\sim$30\%),
while for Group B,  \FeOPT\ FWHM is broader than \MgII\ by 0.1 dex ($\sim$20\%). 

In the case of UV \FeII, we see no significant difference between Group A and B. The FWHM of \FeUV\ and broad emission lines are well correlated 
(with r = 0.73, and r = 0.68, respectively for \Hb\ and \MgII).
On average \FeUV\ is narrower than both \Hb\ (by 0.16 dex) and \MgII\ (by 0.08 dex), indicating that the \FeUV\ emission line region is further out compared to that of these broad emission lines.

\subsection{The \FeOPT\ to \FeUV\ flux ratio}\label{section:flux}

In this section, we examine the flux ratio between \FeOPT\ and \FeUV. The \FeII\ flux is measured in the limited spectral range as described in Section 3.5. We find a large range of the flux ratios from -1.0 to 0.5 with a mean log (\FeOPT/\FeUV) = -0.10 $\pm$ 0.28, indicating complex nature of \FeII emission. 

\begin{figure*}
\figurenum{8}
\centering
	\includegraphics[width = 0.39\textwidth]{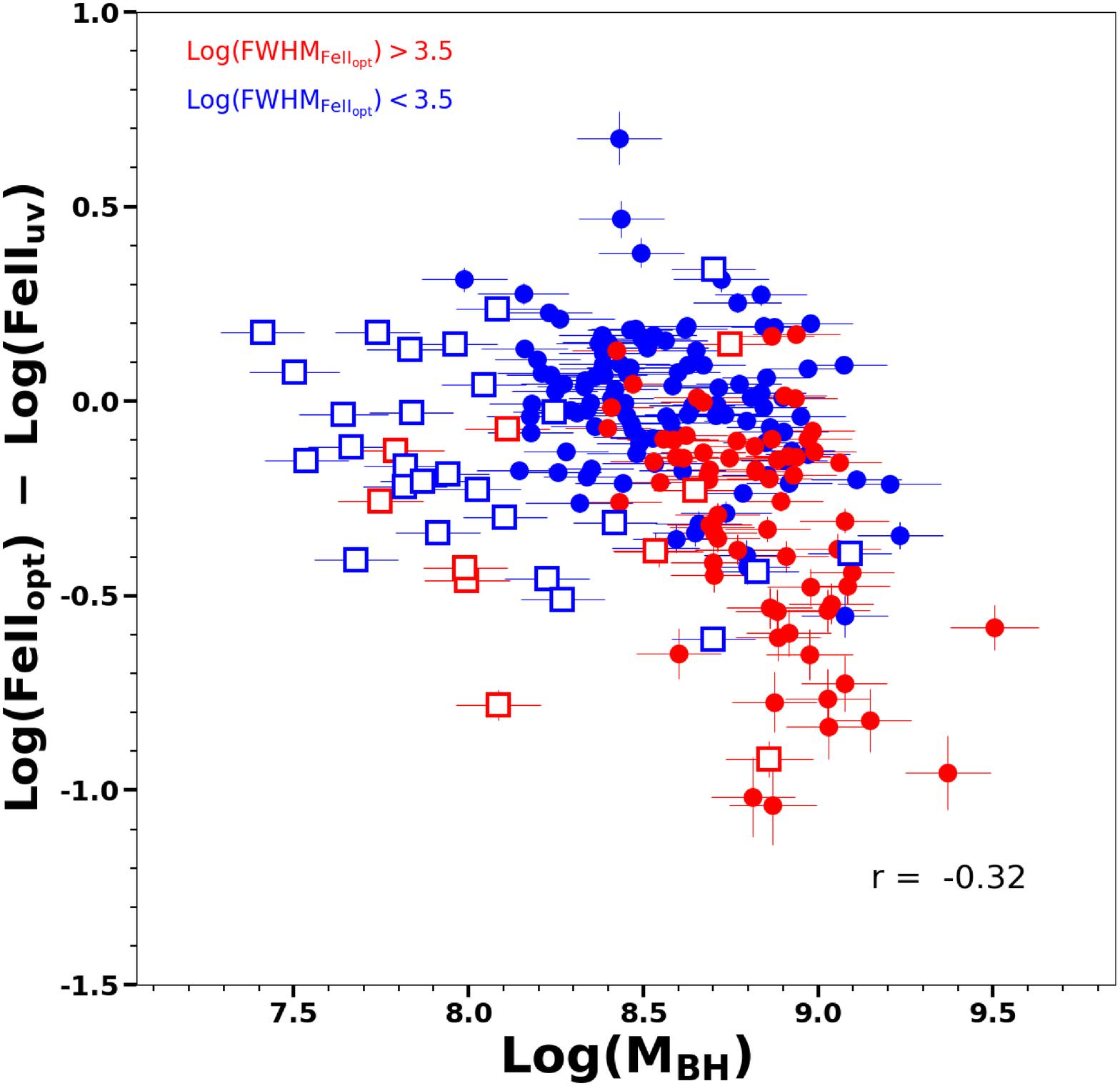}
	\includegraphics[width = 0.455\textwidth]{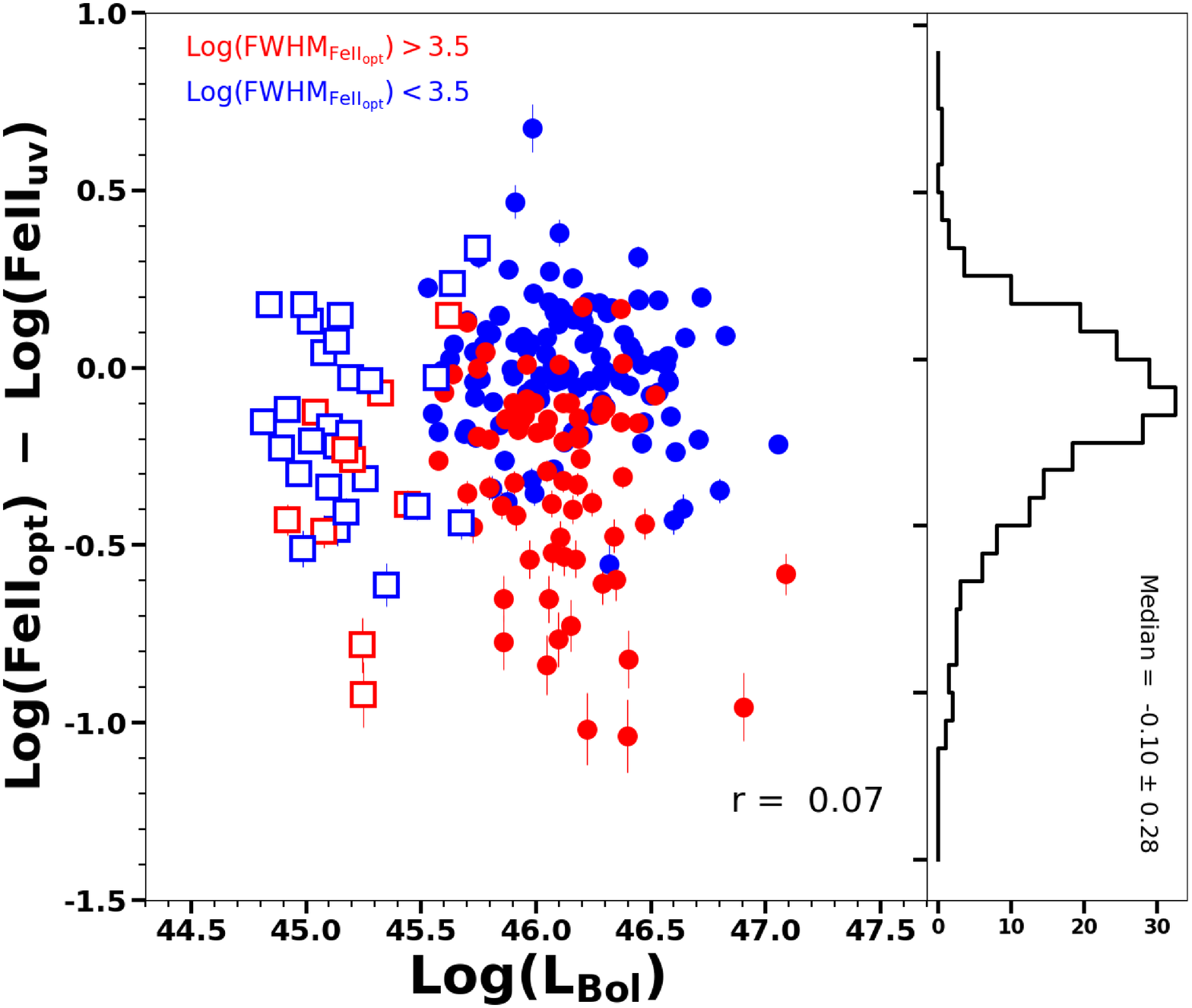}
	\includegraphics[width = 0.42\textwidth]{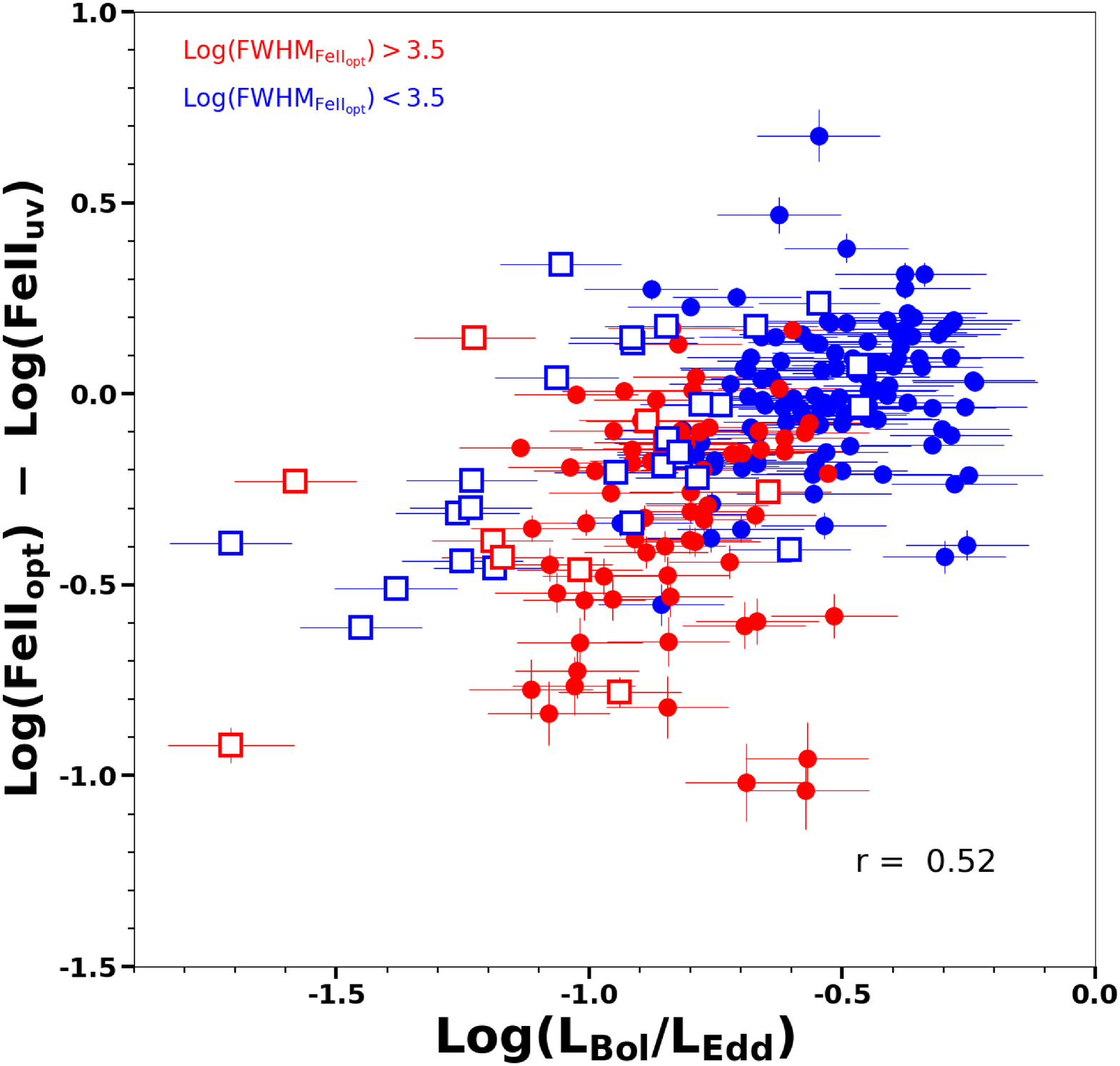}
	\includegraphics[width = 0.42\textwidth]{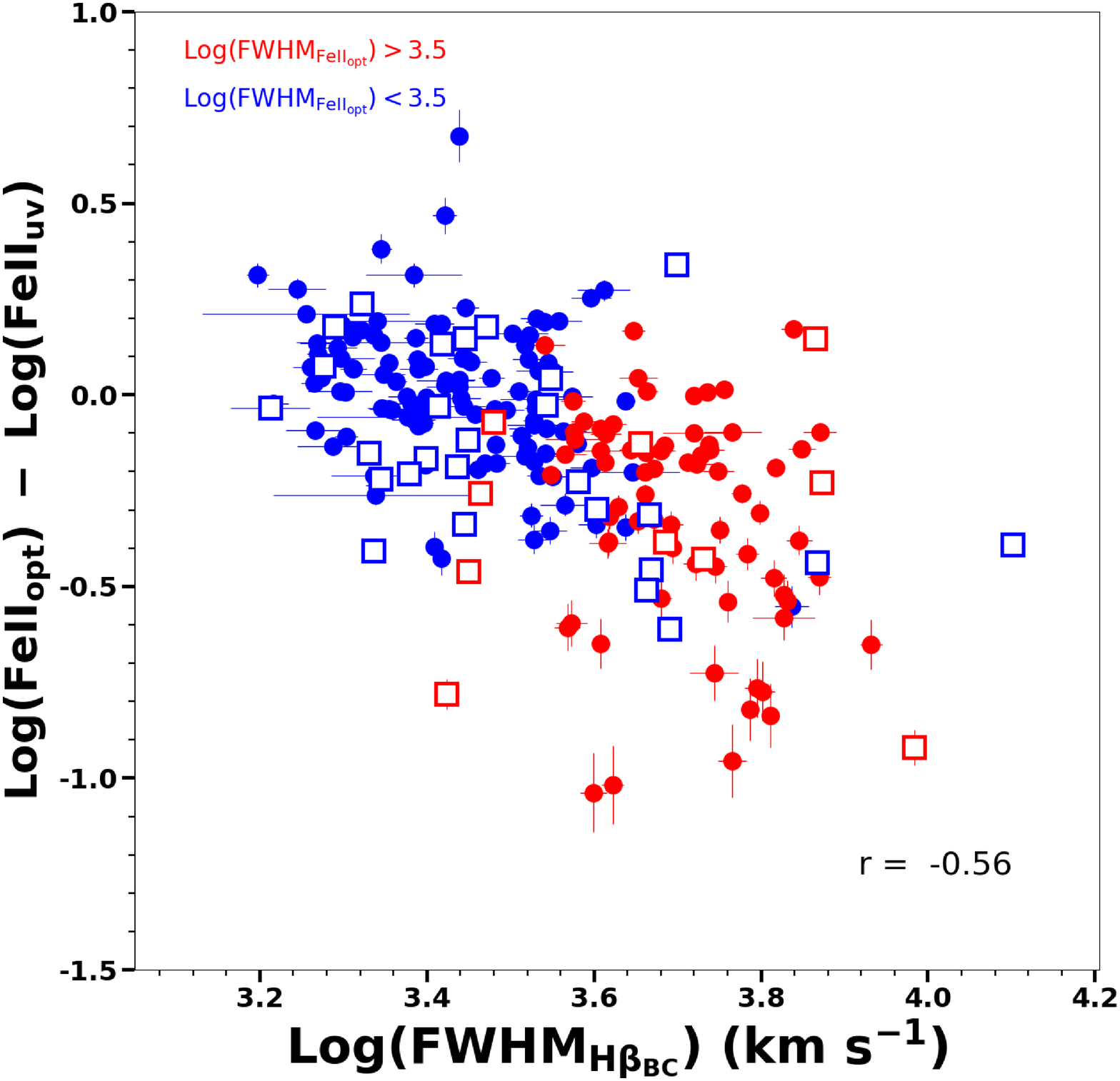}
	\caption{Comparison of the UV and Optical \FeII\ emission line fluxes as a function of AGNs properties. Group A ($\rm FWHM_{FeII_{opt}} < 10^{3.5}$) is denoted with blue symbols, while Group B ($\rm FWHM_{FeII_{opt}} > 10^{3.5}$) is dented with red symbols. Keck and SDSS objects are presented with open and filled symbols, respectively. The correlated coefficients are shown in the lower-left in each panel.}
	\label{fig:flux_uvopt}
\endcenter
\end{figure*}

To investigate whether AGN parameters are related with the optical-to-UV \FeII\ flux ratio, we compare the flux ratio with \mbh, bolometric luminosity, Eddington ratio and the FWHM of \Hb\ in Figure \ref{fig:flux_uvopt}. 
While we found no significant correlation between the flux ratio and bolometric luminosity, the flux ratio shows a weak negative correlation with \mbh\ (i.e., r = -0.32), and a positive correlation with Eddington ratio (r = 0.52). Also, the flux ratio anti-correlates with the line width 
of \Hb\ (r = -0.56). These results are consistent with previous studies \citep[e.g.,][]{Dong+11, Sameshima+11, Kovacevic+15}. Using SDSS sample of 4178 targets, for example, \citet{Dong+11} reported a moderate correlation between \FeOPT$/$\FeUV\ and Eddington ratio, hypothesizing that Eddington ratio is the main driver of controlling the \FeII\ emission line strength since it regulates the distribution of hydrogen density of the emission regions. 
A high Eddington ratio is related to high column density, because when Eddington ratio is high, large radiative pressure could push away low density cloud, thus only high column density gas is gravitationally bound (see Section 3.1 in  \citealp{Dong+11}; \citealp{Sameshima+11}). Similar to our result, \citet{Kovacevic+15} found a negative correlation between \FeII\ flux ratio with FWHM of \Hb, which can be also interpreted similarly since larger \Hb\ line width indicates larger black hole mass and smaller Eddington ratio for a fixed luminosity. 

Differences in the \FeOPT$/$\FeUV\ flux ratio may be caused by differences in the distribution of cloud column densities within the BLR (see Figure 3 in
\citealp{Ferland+09}). \citet{Joly87} showed that \FeOPT\ has smaller optical depth compared to that of \FeUV. Therefore, as the column density increases, \FeOPT\ flux will become larger than that of \FeUV. As a result, the flux ratio of \FeOPT$/$\FeUV\ increases. Thus, AGNs in Group A with on average higher Eddington ratio may have higher column density, which lead to higher \FeOPT$/$\FeUV\ flux ratio.

\section{Discussion}\label{section:discuss}

\subsection{Confirmation of systemic velocity shift of the UV and Optical \FeII\ Emission Lines}

One of our main goals in this paper is to confirm the systemic redshift of the optical \FeII\ emission lines. As we mentioned in Section 1, \citet{Sulentic+12} and \citet{Hu+12} found contrary results on the velocity shift of \FeOPT\ emission using a set of composited spectra. \citet{Sulentic+12} generated high S/N composite spectra using the SDSS AGNs, to test the measurement of \citet{Hu+08}. The spectra were binned based on \FeII\ strengths with a limited range of \Hb\ FWHM $\leq$ 4000 \kms\ or  \Hb\ FWHM between $4000$ and $8000$ \kms. The composite spectra with 
high S/N ($\sim$55$-$60) composite spectra did not show a systemic redshift of optical \FeII\ with respect to \OIII. However, by combining AGNs with the simialr $\rm V_{FeII}$, \citet{Hu+12} argued that their measurement of the systemic redshift was reliable.
While these two studies used composite spectra with a very high S/N, which were binned with different criteria, the results lead to discrepancy. 

The solution to this issue is to use individual spectra with a high S/N, and we were able to measure the velocity shift of \FeOPT\ using a large sample of luminous AGNs with high S/N spectra. In Section \ref{section:confirm}, we present the distribution of \FeOPT\ (Figure \ref{fig:kinematic_uvopt}), showing that \FeOPT\ show a large range of velocity shift with an average $\sim$182 $\pm$ 95 \kms. Our results confirmed that optical \FeII\ emission lines are
redshifted on average. The discrepancy between \citet{Hu+08} and \citet{Sulentic+12} may be caused by the different criteria in constructing composite spectra. \citet{Sulentic+12} combined the spectra over the limited ranges of FWHM of \Hb\ and \FeII\ strength, while \citet{Hu+12} combined the spectra based on the limited ranges of \FeII\ velocity. \citet{Hu+12} argued that \citet{Sulentic+12} failed to find the systemic redshift of \FeOPT\ because they made composite spectra using similar FWHM \Hb\ and \FeII\ strength, averaging the velocity shift of individual objects.

We also find the average velocity shift of \FeUV\ as 40 $\pm$ 141 \kms, which is smaller than that of the previous measurement, 1150 $\pm$ 580 \kms\ by \citet{Kovacevic+15}. Note that \citet{Kovacevic+15} mentioned that their measured velocity shift of \FeUV\ should be taken into caution because of large uncertainty for individual AGNs with very broad UV \FeII\ emission, while it is not clear why the measured average velocity shift of \FeUV\ is different between \citet{Kovacevic+15} and ours. One possibility is that the difference may be caused by different \FeII\ templates. In our analysis, we adopted the I Zw 1 \FeII\ template of \citet{Tsuzuki06}, while \citet{Kovacevic+15} used the UV \FeII\ template from \citet{Kovacevic+10}. Recently, \citet{Shin+19} showed that the flux ratio of \FeII/\MgII\ could be different up to $\sim$0.2 dex if the flux of \FeII\ is modeled from different \FeII\ templates such as \citet{Vestergaard+01} and \citet{Tsuzuki06}. 

\subsection{Origins of the UV and Optical \FeII\ Emission Lines}\label{section:connection_ki}

We showed that the widths of \FeUV\ and \FeOPT\  lines are comparable, suggesting that the two emission line regions are close to each other, as reported and suggested by previous studies \citep{Hu+08, Kovacevic+15}. Interestingly, AGNs in Group A and Group B show a different trend in comparing the widths of \FeUV\ and \FeOPT. For AGNs with a relatively narrow \MgII\ line (i.e., $<$ 4000 \kms; Group A), the widths of \FeUV\ and \FeOPT\ are comparable (Figure \ref{fig:fwhm_uvopt}).
In contrast, for AGNs with a very broad \MgII\ line (i.e., $>$ 4000 \kms; Group B), \FeOPT\ is broader than \FeUV. Similarly, the velocity shifts of \FeUV\ and \FeOPT\ are better correlated in Group A, while there is a much larger scatter in Group B. 
These results suggest that the \FeII\ emission region is more complex for the AGNs with large gas velocities. Previously, \citet{Kovacevic+15} reported a large average velocity shift of \FeUV, but no significant velocity shift of \FeOPT. \citet{Kovacevic+15} discussed that the UV and optical \FeII\ emission lines may be located in the same region, however, if \FeOPT\-emitting gas clouds are more symmetrically distributed than UV-emitting gas, potentially causing the difference in the velocity shift. In addition, \citet{Kovacevic+15} discussed the effect of internal shock waves in the infalling gas. As the shock may contribute the excitation of UV lines, a larger redshift is expected in \FeUV. 

Following the study of \citet{Ferland+09}, who focused on the anisotropic properties of \FeII\ emission based on a photoionization calculation 
for a single cloud in the BLR,  we consider a couple of effects on the \FeII\ line emission.  
First, we expect anisotropy in \FeII\ emission due to the various contribution from the front illuminated side and the back shielded face of each cloud. It is expected that the anisotropy is stronger for higher column density clouds, and that \FeUV\ emission is emitted more asymmetric since UV \FeII\ emission arises from larger optical thick transitions compare to that of optical \FeII\ emission. In contrast, \FeOPT\ emission is expected to be more isotropic with symmetry between illuminated and shielded sides of individual clouds.  
Second, the observed \FeII\ emission is the combination of the flux from individual clouds with varying column density and distance from the central 
photoionizing source. As shown by \citet{Ferland+09}, clouds with higher column density are expected to have less acceleration due to radiation pressure and show infall signature, i.e., redshifted \FeII\ emission. 
Thus, depending on the column density distribution of individual clouds in the BLR, the observed \FeII\ emission may show different signatures in the velocity shift and line width between the UV and optical \FeII\ emission lines.

As shown in Figure 8, AGNs in Group A have higher Eddington ratio and higher optical-to-UV \FeII\ flux ratio than AGNs in group B, indicating that the average column density of individual clouds is higher. Thus, we may expect stronger anisotropy in \FeII\ emission from individual clouds. Nevertheless, 
we find no corresponding kinematical signature in Group A in comparison with Group B, since the relative velocity shift between optical and UV \FeII\ emission is similar between the two groups (see Figure 6). The scatter of the relative velocity shift is even smaller in Group A than in Group B. In the case of the line width, we also see that AGNs in Group A show similar line widths between the optical and UV \FeII\ emission. These results may imply that the observed velocity shift and line width, which are flux-weighted over a large number of individual clouds, may not be directly influenced by the anisotropy of \FeII\ emission from each cloud. 

While the model of \citet{Ferland+09} has successfully explained the redshift of \FeII\ (e.g., \citealp{Hu+08}, \citealp{Kovacevic+15}), 
and supports our result that the UV and optical \FeII\ emission lines are on average redshifted (40 $\pm$ 141 \kms\ and 182 $\pm$ 95, respectively for \FeUV\ and \FeOPT, with respect to the peak of the \Hb\ line), there are also AGNs with blueshifted \FeII, particularly in UV emission. These AGNs may indicate outflows of gas clouds. Note that each cloud in the BLR can be accelerated outward due to strong radiation pressure or infall due to weak acceleration from radiation pressure, depending on the column density of each cloud compared to the minimum column density required for infall (see Equation 2 in \citealp{Ferland+09}). Thus, for a given AGN, there could be a mix of outflows and infall in the BLR. Without spatially-resolving individual clouds, we only see the average velocity shift of flux-weighted \FeII\ emission. This scenario may explain our result, which shows a large range of average velocity shift of \FeII\ (e.g., $\pm$1500 \kms).

\section{Conclusions}\label{section:sum}

In this paper, we investigate the kinematical properties of the UV and optical \FeII\ emission lines, using a sample of 223 AGNs at 0.4 $<$ z $<$ 0.8 with a high S/N ratio in the continuum, which allowed us to explore the connection between the UV and optical \FeII\ emission regions. The main results are summarized as follows:

\smallskip
(1) We find a strong correlation between the widths of UV and optical \FeII\ emission lines, supporting that both UV and optical \FeII\ emissions arise from an approximately same distance in the BLR. However, we find a different trend depending on the width of \FeII. The line widths of UV and optical \FeII\ emission lines are comparable to each other for AGNs with a relatively small \FeOPT\ line width (i.e., FWHM $<$ 3200 \kms; Group A), while for AGNs with a very broad \FeOPT\ (i.e., FWHM $>$ 3200 \kms; Group B), \FeOPT\ is broader than \FeUV. 

\smallskip
(2) \FeII\ emission lines are on average narrower than \Hb\ and \MgII\ for Group A, indicating the \FeII\ emission region is further out in the BLR. This result is consistent with the previous study by \citet{Hu+08} for the optical \FeII\ emission. However, for AGNs with very broad \FeII, \FeOPT\ emission line is comparable to or broader than \Hb\ or \MgII. 

\smallskip
(3) We confirm the systemic redshift of optical \FeII\ emission lines with an average velocity 182 $\pm$ 95 \kms, as similarly reported by \citet{Hu+08} and \citet{Kovacevic+15}, while for individual AGNs there is a large range of the velocity shift, including blueshift. In the case of UV \FeII\ emission lines, we find the average velocity 40 $\pm$ 141 \kms. The average velocity shift of the UV and optical \FeII\ emission may indicate inflow motion. However, there are AGNs with various signs of the velocity shift, indicating complex nature of \FeII\ emission. 

\acknowledgements

We thank the anonymous referee for various comments and suggestions, which improved the clarity of the paper. This work has been supported by the Basic Science Research Program through the National Research Foundation of Korea government (2016R1A2B3011457 and No.2017R1A5A1070354). 
We thank Prof. Yongquan Xue for helpful discussions. H.A.N.L. and Y.Q.X. acknowledge support from the Chinese Academy of Sciences President's International Fellowship Initiative. Grant No. 2019PM0020.
\\


\end{document}